\documentclass[11pt, a4paper]{article}
\pdfoutput=1

\usepackage{jcappub}
\usepackage{graphicx}
\usepackage{slashed}
\usepackage{booktabs}
\usepackage{multirow}
\usepackage{color}
\usepackage{textpos}

\title{Galactic dark matter search via phenomenological astrophysics modeling}

\author[1]{Xiaoyuan Huang,}
\author[2,3,4]{Torsten En{\ss}lin,}
\author[2,5]{Marco Selig}

\affiliation[1]{Physik-Department T30d, Technische Universit\"at M\"unchen, James-Franck-Stra{\ss}e, D-85748 Garching, Germany} 
\affiliation[2]{Max-Planck-Institut f\"ur Astrophysik, Karl-Schwarzschildstra{\ss}e
1, D-85748 Garching, Germany}
\affiliation[3]{Ludwig-Maximilians-Universit\"at M\"unchen, Geschwister-Scholl-Platz
1, D-80539 Munich, Germany}
\affiliation[4]{Exzellenzcluster Universe, Technische Universit\"at M\"unchen, Boltzmannstr.
2, D-85748 Garching, Germany}
\affiliation[5]{IBM R\&D GmbH, Sch\"onaicher Stra\ss{}e 220, D-71032 B\"oblingen, Germany}
\emailAdd{xiaoyuan.huang@tum.de}
\emailAdd{ensslin@mpa-garching.mpg.de}
\emailAdd{mselig@mpa-garching.mpg.de}

\abstract{Previous searches for the $\gamma$-ray signatures of annihilating
galactic dark matter used predefined spatial templates to describe
the background of $\gamma$-ray emission from astrophysical processes
like cosmic ray interactions. In this work, we aim to establish an
alternative approach, in which the astrophysical components are identified
solely by their spectral and morphological properties. To this end,
we adopt the recent reconstruction of the diffuse $\gamma$-ray sky
from Fermi data by the D$^{3}$PO algorithm and the fact that more
than 90\% of its flux can be represented by only two spectral components,
resulting form the dense and dilute interstellar medium. Under these
presumptions, we confirm the reported DM annihilation-like signal
in the inner Galaxy and derive upper limits for dark matter annihilation
cross sections. We investigate whether the DM signal could be a residual
of the simplified modeling of astrophysical emission by inspecting
the morphology of the regions, which favor a dark matter component.
The central galactic region favors strongest for such a component
with the expected spherically symmetric and radially declining profile.
However, astrophysical structures, in particular sky regions
which seem to host most of the dilute interstellar medium, obviously would
benefit from a DM annihilation-like component as well. 
Although these regions do not drive the fit, they warn that a more detailed understanding
of astrophysical $\gamma$-ray emitting processes in the galactic
center region are necessary before definite claims about a DM annihilation
signal can be made. 
The regions off the Galactic plane actually  disfavor the best fit DM annihilation cross section from the inner Galactic region unless the radial decline of the Galactic DM density profile in the outer regions is significantly steeper than that usually assumed. }

\begin{document}
\begin{textblock}{3}(9.3,0)
  \noindent
  TUM-HEP  1029/15
\end{textblock}

\maketitle

\section{Introduction}
The presence and therefore the existence of dark matter (DM) in the
Universe has been confirmed through many independent astrophysical
studies, while the physical nature of DM particles is still unknown
\citep{Zwicky:1933gu,Jungman:1995df,Bertone:2004pz,Ade:2013zuv}.
A promising way to probe the nature of DM is to identify its annihilation signatures. In many scenarios, DM particles froze-out in the primordial Universe, and this kind of thermal production promises a non-negligible annihilation cross section. The dense centers of galaxies and galaxy
clusters should therefore be sites of DM annihilation events. The
thereby produced charged particles and photons might be detectable
and thereby reveal particle properties of DM \citep{Adriani:2008zr,Aguilar:2013qda,Adriani:2011xv,Chang:2008aa,FermiLAT:2011ab,Atwood:2007ra,Atwood:2009ez}.
Unlike any charged particle created by DM annihilation, which looses
its directional information during its propagation, annihilation $\gamma$-rays should provide precise information on
the spatial distribution of DM as well as direct hints on the mass
of DM particles in some extreme cases \citep{Bringmann:2012vr,Weniger:2012tx,Ackermann:2013uma,Ackermann:2015lka,Ibarra:2015tya,Huang:2015fca}.
Sky locations such as dwarf galaxies, galaxy clusters and the Galactic
center, which are promising for indirect detection of DM using $\gamma$-ray
data, should contain DM in high density, be relatively nearby, and
show little flux of astrophysical (not-DM-annihilation related) $\gamma$-rays.
The GC region is ideal with respect to the first two conditions, however,
it exhibits unfortunately significant astrophysical $\gamma$-ray
sources as it harbors supernovae explosions injecting cosmic rays
(CRs) into the interstellar medium (ISM) and compact sources of high
energy particles and radiation \citep{vanEldik:2015qla}. 

Several groups have analyzed the publicly-available Fermi-LAT data
and reported a spatially extended $\gamma$-ray excess
of around $1-3$  GeV from the region surrounding the Galactic
Center (GC) with respect to the expected diffuse Galactic $\gamma$-ray
emission (DGE) of astrophysical, non-DM origin \citep{Goodenough:2009gk,Vitale:2009hr,Hooper:2010mq,Hooper:2010im,Abazajian:2012pn,Gordon:2013vta,Huang:2013pda,Hooper:2013rwa,Daylan:2014rsa}.
In general, it was found that dark matter with several tens of GeV,
which annihilates into $b\bar{b}$ and $\tau^{-}\tau^{+}$ final states,
would account for the spectral shape \citep{Calore:2014xka,Agrawal:2014oha}
and a generalized NFW profile \citep{Navarro:1995iw,Navarro:1996gj}
with an inner slope $\alpha$ = 1.2 would explain the spatial extension
of this excess \citep{Hooper:2013rwa,Daylan:2014rsa,Calore:2014xka}.

The DGE model assumed in these works was usually that of the Fermi
collaboration. This model is constructed by assuming the flux to be
described by a linear superposition of spatial templates, which are
partly directly observations at other wavelength, partly the result
of physical modeling of the CR propagation and interaction in the
Galaxy. 

It is anticipated that the DGE model uncertainties affect the apparent
GeV $\gamma$-ray excess towards the GC, which implies considerable
systematic uncertainties for the deduced DM properties or upper limits.
Zhou et al. \cite{Zhou:2014lva} and Calore et al. \cite{Calore:2014xka} address this by
scanning the model parameter space by changing the simulated CR source
distributions, CR propagation parameters, and CR target properties.
The presence of the GeV excess signal seems to be robust with respect
to the resulting changes in the DGE template, however, the spectral
shape of the GeV excess seems to vary to some degree. 

In addition to their investigation of theoretical uncertainties of
the different DGE models, Calore et al. \cite{Calore:2014xka} performed a principal
component analysis of the $\gamma$-ray flux residuals in a number
of test regions along the Galactic plane which permitted them to estimate
empirically their model uncertainties. The best fit DM parameters
derived in this careful investigation will be compared to the results
presented from our analysis.  

Apart from DM annihilation, $\gamma$-ray emission from millisecond
pulsars could also be the origin of the GeV excess. Millisecond pulsars
are believed to be abundant in the GC and their spectra also peak
at several GeV. A blend of unresolved millisecond pulsars should appear
as extended $\gamma$-ray emission to the Fermi-LAT instrument \citep{Abazajian:2012pn,Gordon:2013vta,Mirabal:2013rba,Yuan:2014rca,Calore:2014oga}.
However no millisecond pulsar has yet been identified in the GC region.
This non-detection questions the millisecond pulsar scenario \citep{Hooper:2013nhl,Cholis:2014lta}
unless their brightness distribution function resides mostly below
the sensitivity of Fermi-LAT \citep{Petrovic:2014xra}. Recent investigations
\citep{2015arXiv150605124L,2015arXiv150605104B} show that unresolved
point sources could indeed account for the GeV excess. 

The unclear picture drawn by these studies motivates us to scrutinize
the reported potential DM annihilation signal for the possibility
that the reported $\gamma$-ray excess is due to imperfections in
the DGE and point source modeling. To this end, an alternative, template-free,
non-parametric, and phenomenological DGE and point source model, which
significantly differs from that of the Fermi collaboration and other
groups, is used in the following search for a Galactic $\gamma$-ray
DM signal.

\section{The Gamma-ray sky}
\label{sec:sky}
\subsection{The astrophysical gamma ray sky}
Our DGE model is based on the recent re-analysis of the Fermi $\gamma$-ray
data in terms of diffuse flux and point sources by Selig et al. \cite{Selig:2014qqa}.
Selig et al. \cite{Selig:2014qqa} used the D$^{3}$PO algorithm \citep{2015A&A...574A..74S}
to produce noise- and point source-cleaned maps of the diffuse $\gamma$-ray
emission at nine logarithmically spaced energy bands ranging from
just below 1 GeV to 300 GeV. 

D$^{3}$PO stands for ``Denoising, deconvolving, and decomposing
photon observations'' and is an imaging algorithms for high energy
photons. D$^{3}$PO assumes the $\gamma$-ray sky to be composed of
a diffuse component, which varies on a logarithmic scale but exhibits
spatial correlations, and a point-source component, which consists
of spatially uncorrelated point sources with a power-law brightness
distribution. Using this, D$^{3}$PO decomposes the observed photon
flux into these components probabilistically while also taking the
instrument response and the Poisson statistics of the $\gamma$-ray
events into account. The D$^{3}$PO algorithm is derived in the context
of information field theory \citep{Ensslin:2008iu}, the information
theory for fields, and implemented using the numerical information
field theory (NIFTY) library \citep{Selig:2013ta}, which permits
the construction of space-pixelization independent code. In consequence,
D$^{3}$PO provides two sky maps, one containing the diffuse emission
component and one the point source component, the latter having virtually
a source in each pixel, however, most of them being infinitesimally
weak. 

In contrast, astrophysical gamma ray model based on the Fermi Collaboration
data products account only for the presence of the point sources which
were detected above the significance threshold chosen by the Fermi
Collaboration. This permits the possibility that dense sub-threshold
sources appear as an extended source, distinct from the astrophysical
$\gamma$-ray model, which might mimic DM annihilation.

In contrast to this, the analysis by Selig et al. \cite{Selig:2014qqa}
takes marginally detected point sources into account probabilistically.
Only populations of even weaker, individually unresolved point sources
would be missed by D$^{3}$PO as well. Whether such sources could
account for the GeV excess is an open question \citep{2015arXiv150605104B,2015arXiv150605124L}.

An important observation of Selig et al. \cite{Selig:2014qqa} was that
more than 90\% of the DGE at all sky locations and all investigated
energies could be accounted for by a simple, phenomenologically constructed
two component model: The $\gamma$-ray spectra of a molecular cloud
complex in the Galactic anti-center and that of the southern tip of
the southern Fermi bubble \citep{ Dobler:2009xz,Su:2010qj,Fermi-LAT:2014sfa} served
as spectral templates in a point-to-point spectral fitting of the
nine D$^{3}$PO maps at different energies. The result of this analysis
showed ``cloud-like'' gamma-ray emission with a spatial morphology
closely resembling that of Galactic dust maps derived from completely
independent microwave \citep{Abergel:2013fza} or far-infrared \citep{Finkbeiner:1999aq}
observations. This, as well as the steep spectrum with a hint of a
bump at GeV as typical for $\gamma$-rays from the decay of $\pi^{0}$s
indicates that the ``cloud-like'' component is mostly due to hadronic
CR proton interactions in the denser parts of the ISM \citep{Selig:2014qqa}.
The ``bubble-like'' component showed prominently the Fermi-bubbles,
as well as an puffed up version of the Galactic disc. This morphology,
as well as the harder, power-law like spectrum indicates that this
component is mostly due to inverse Compton up-scattering of photons
by CR electrons, which predominantly occurs in the volume-dominating
hot and dilute part of the ISM \citep{Selig:2014qqa}. The Fermi
bubbles are then just two giant outflows driven by the hot ISM and
in particular the light CR content of it, as a number of authors already
suggested \citep{Su:2010qj,Carretti:2013sc}.

The association of the two radiation processes to the two ISM phases
is certainly not unique, as each of them will happen in each phase,
but the preferences are certainly correct due to the very different
nuclear target densities. 

Since the phenomenological two component description of the DGE successfully
captures the dominant $\gamma$-ray properties of the Milky Way, we
will assume it to be correct within this work. As both used spectral
templates were taken from regions far from the GC, they should be
little contaminated by $\gamma$-rays from potential DM annihilation
or other processes only predominantly occurring in the GC. Any such
contribution should -- if it does not mimic our spectral templates
-- be visible in terms of excess $\gamma$-rays with respect to our
two component model. In the following we will test the data for such
excess photons. 

Our phenomenological two component DGE model can certainly be criticized
for its simplicity, the neglect of spectral differentiating CR transport
processes, and its ignorance to the existing detailed knowledge on
radiation processes. It is meant as an orthogonal approach to the
existing DM searches, which should show which aspects of these investigations
are robust, and which might need improvements. We believe that an
unmatched strength of our approach is the flexible, non-parametric
form of the spatial template generation. In the approaches so far,
the used, more rigid spatial templates could potentially be inadequate
to represent the real DGE.

Our spectral resolution will be limited to the energy bin choise of Selig et al. \citep{Selig:2014qqa}. Although a higher spectral resolution than nine energy bands in the range of $\sim 1\ldots 300$ GeV would certainly be desirable, this is not feasible. The current version of the D$^3$PO algorithm has to process individual bands separately and requires that each band has a sufficiently high photon statistics to perform a good decomposition of the sky into point sources and diffuse flux. As shown below, our analysis reproduces well the reported GC $\gamma$-ray excess and finds very similar DM parameters as reported in Calore et al. \cite{Calore:2014xka}. This indicates that the limited spectral resolution of our analysis does not hamper our sensitivity for DM annihilation.

\subsection{DM annihilation}

Following previous work \citep{Hooper:2013rwa,Daylan:2014rsa,Calore:2014xka},
we model the radial distribution of Galactic DM 

\begin{equation}
\rho(r)=\frac{\rho_{s}}{(r/r_{s})^{\alpha}(1+r/r_{s})^{3-\alpha}}
\end{equation}
as a generalized NFW profile \citep{Navarro:1995iw,Navarro:1996gj}
with an inner slope of $\alpha$ = 1.2. We determine the normalization
$\rho_{s}$ by fixing the DM density at the solar radius to $\ensuremath{\rho(r_{\odot}=8.5}\ kpc\ensuremath{)}=0.4\ \mbox{GeV cm}\ensuremath{^{-3}}$ \cite{Catena:2009mf,Salucci:2010qr,Iocco:2011jz}. 

We investigate the DM parameter space spanned by the DM mass $m_{dm}$
and velocity-averaged cross sections $\langle\sigma v\rangle$. We
scan through this parameter plane in a logarithmically equal spaced
grid with $m_{dm}$ taking twenty values between $5\,\mbox{GeV}/c^{2}$
and $200\,\mbox{GeV}/c^{2}$ as well as $\langle\sigma v\rangle$
taking fifty values in between $10^{-5}$ to 5 in unit of $3\times10^{-26}\ \mathrm{cm}^{3}\mathrm{s}^{-1}$.
As in previous works, we investigate the most common annihilation
final states $b\bar{b}$ and $\tau^{-}\tau^{+}$, and we use the corresponding
spectrum derived from PPPC4DMID, in which electro-weak corrections are included
\citep{Cirelli:2010xx}. 

\subsection{The likelihood\label{sub:The-likelihood}}
To be consistent with the adopted DGE modeling, we use the same data
set as in \citep{Selig:2014qqa}. The 6.5 years Fermi reprocessed
Pass 7 CLEAN data are binned in nine logarithmically spaced energy
bands ranging from 0.6 to 307.2 GeV, and the sky is discretized by
HEALPix scheme with n$_{side}$=128, which corresponds
to an angular resolution of approximately 0.46$^{\circ}$ . For each
energy bin, the data are further split into FRONT and BACK events
according to where in the LAT instrument the photons where registered.
D$^{3}$PO has difficulties to accurately model the $\gamma$-ray
sky in the highest energy bin since the low number of photons there
inhibit a clear separation into diffuse and point like flux. For this
reason, Selig et al. \cite{Selig:2014qqa} ignored this bin in any further
analysis and we will do so as well. The corresponding Fermi LAT instrument response and exposure functions
necessary for the data analysis are generated by Fermi Science Tools\footnote{http://fermi.gsfc.nasa.gov/ssc/data/analysis/software/}. 

As described above, D$^{3}$PO could decompose the $\gamma$-ray sky
into point source components and diffuse component, in which the contribution
of possible DM annihilation could be embedded. The ``cloud-like''
and the ``bubble-like'' components from Selig et al. \cite{Selig:2014qqa}
are convolved with the energy dependent instrument response and exposure
functions yielding the expected number of counts, $n_{c}^{ijk}$and
$n_{b}^{ijk}$ respectively, in each pixel $i$, each energy bin $j$,
and for each photon detection mode $k$ (FRONT or BACK). The combined
index $ijk$ fully indexes our data space here and in all following
formula. 

This model of expected counts is however constructed without taking
the presence of a third component, annihilating DM, into account.
In order to fix this and to release DM annihilation flux potentially
absorbed by the DGE components, their spatial morphology is made flexible
here. To this end, two free fitting parameters $\alpha_{i}$ and $\beta_{i}$
are introduced for each spatial pixel $i$, such that $n_{c}^{ijk}\rightarrow\alpha_{i}n_{c}^{ijk}$
and $n_{b}^{ijk}\rightarrow\beta_{i}n_{b}^{ijk}$. 

For fixed DM properties, such as the annihilation final states, $m_{\mathrm{dm}}$,
and $\langle\sigma v\rangle$, the expected number $n_{\mathrm{dm}}^{ijk}$
of DM induced $\gamma$-ray counts for all data space locations $ijk$
can be calculated. These, plus the corresponding expected photon counts
due to the ``cloud-like'' $\gamma$-ray emission, $\alpha_{i}n_{\mathrm{c}}^{ijk}$
, the ``bubble-like'' $\gamma$-ray emission, $\beta_{i}n_{\mathrm{b}}^{ijk}$,
and the point sources, $n_{\mathrm{point}}^{ijk}$, form the total
expected $\gamma$-ray counts

\begin{equation}
\lambda^{ijk}=n_{dm}^{ijk}+\alpha_{i}n_{c}^{ijk}+\beta_{i}n_{b}^{ijk}+n_{point}^{ijk}.\label{eq:lambdda}
\end{equation}
The expected $\gamma$-ray counts $\lambda^{ijk}$ depend on the assumed
DM model ($m_{\mathrm{dm}}$ and $\langle\sigma v\rangle$) and set
of fitting parameters ($\alpha=(\alpha_{i})_{i}$, $\beta=(\beta_{i})_{i}$),
collectively called the parameters $p=(m_{\mathrm{dm}},\langle\sigma v\rangle,\alpha,\beta)$.
These expected counts should be compared to the actually observed
number of photons $n_{obs}^{ijk}$ to infer these parameters $p$.
We do this by minimizing the objective functions given by the negative
log-likelihood 
\begin{eqnarray}
\chi_{\mathrm{ROI}}^{2}(p) & \equiv & \sum_{i\in\mathrm{ROI}}\chi_{i}^{2}(p)\mbox{ with}\nonumber \\
\chi_{i}^{2}(p) & \equiv & -2\ln L_{i}\left(d_{i}|p\right)
\end{eqnarray}
for any special region of interest (ROI) pixel-wise with respect to
$\alpha_{i}$ and $\beta_{i}$, the re-normalizations of the ``cloud-like''
and ``bubble-like'' $\gamma$-ray emission, while scanning through
the DM parameter subspace. Here, the data vector $d_{i}=\left(n_{\mathrm{obs}}^{ijk}\right)_{jk}$
of the counts associated with pixel $i$ is introduced as well as
the the pixel-wise Poisson count likelihood $L_{i}\left(d_{i}|p\right)$,
by
\begin{equation}
\ln L_{i}\left(d_{i}|p\right)=\sum_{jk}\left[n_{\mathrm{obs}}^{ijk}\ln\lambda^{ijk}-\lambda^{ijk}-\ln\left(n_{\mathrm{obs}}^{ijk}!\right)\right].
\end{equation}
The last term in the brackets has no influence on the fit and is therefore
ignored in the analysis.

\begin{figure*}[!htb]
\centering
\includegraphics[width=0.48\textwidth]{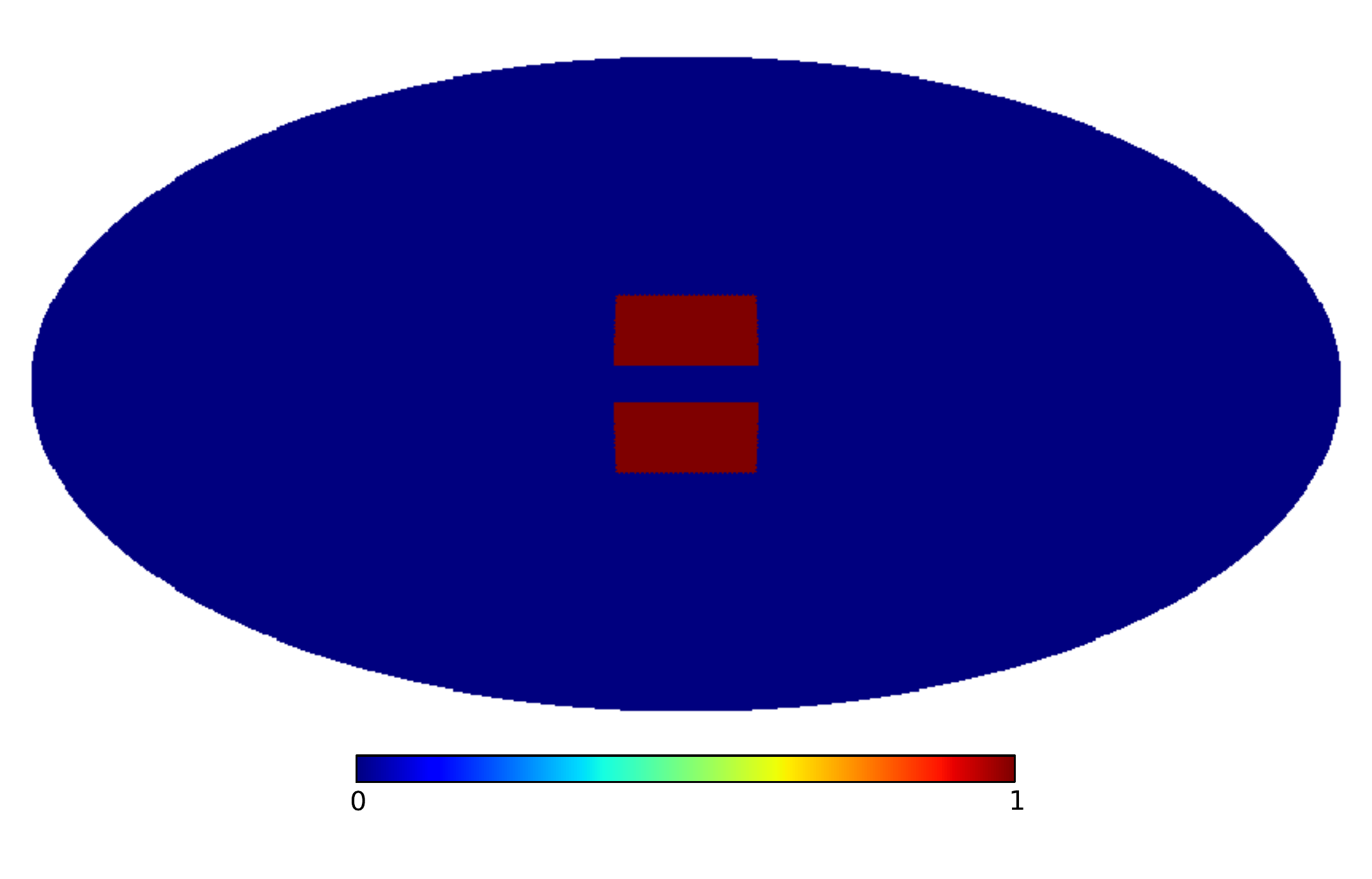}
\includegraphics[width=0.48\textwidth]{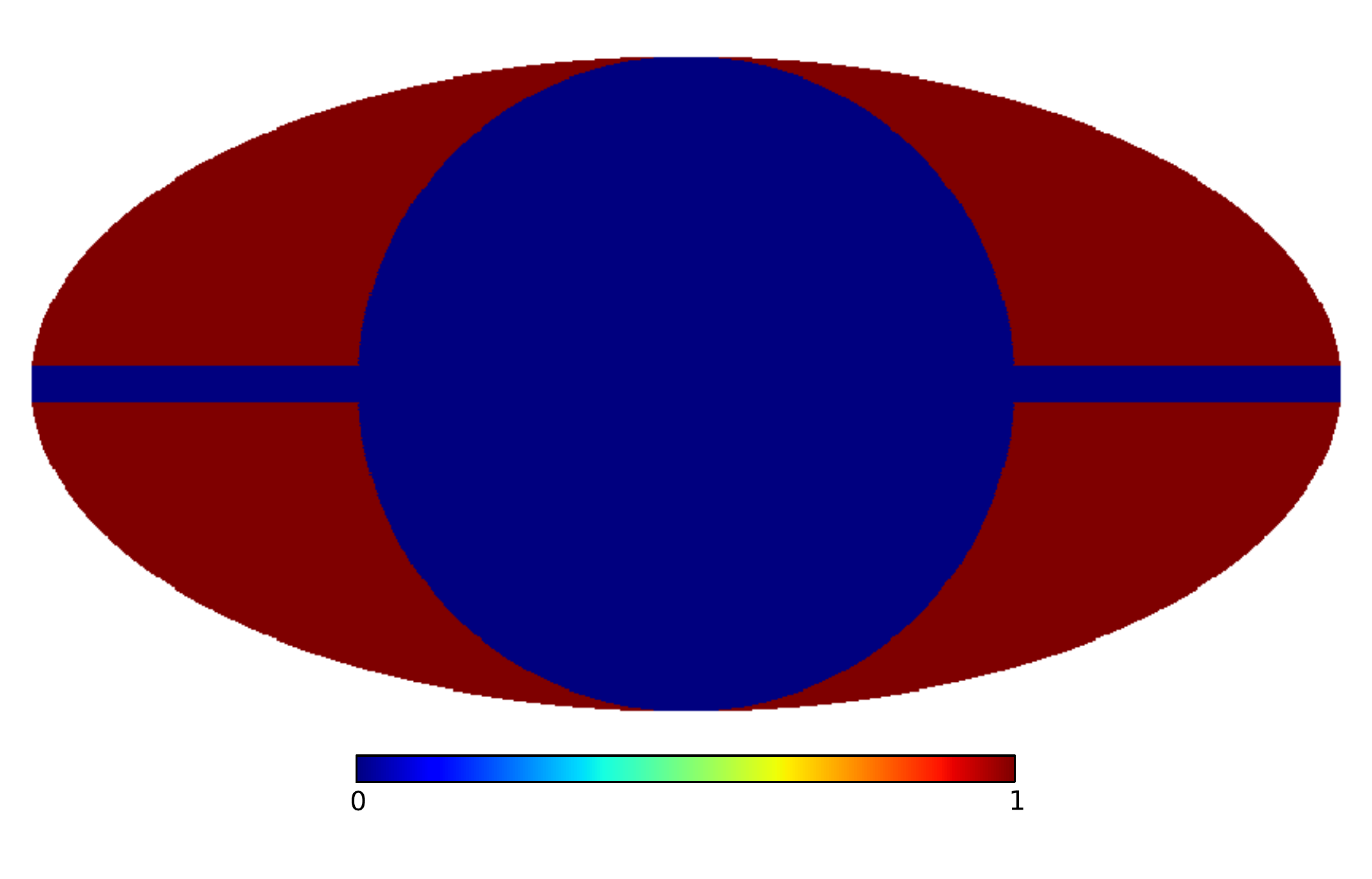}
\caption{Regions used in our analysis. The primary ROI is left and the test
ROI is right.}
\label{fig:ROI}
\end{figure*}

The DM distribution is concentrated towards the center of the Milky
Way, which is, unfortunately, one of the most complicated sky area
due to the various astrophysical sources there. To become insensitive
to our probably imperfect modeling of the central Galactic region,
we define ROIs which exclude it from our analysis. Furthermore, since
numerous faint, undetected and therefore not-modeled point sources
may reside in the Galactic plane, which nevertheless might contaminate
the diffuse emission, we also mask the Galactic plane for these ROIs
to ensure the validity of our two components astrophysical diffuse
model. As our primary ROI we select
Galactic latitudes $4{^\circ}<|b|<20{^\circ}$ and Galactic longitudes
$|l|<20{^\circ}$ similar to the ROI used in  \citep{Calore:2014xka},
but masking a bit more of the Galactic plane region. For comparison,
we select also a test ROI, which excludes any area close
to the GC and close to the galactic plane. For the test ROI we expect
the contribution of DM annihilation to the $\gamma$-ray sky to be
negligible. These ROIs are shown in Fig.~\ref{fig:ROI}.

\section{Results}

\subsection{Pure astrophysics}

\begin{figure*}[!htb]
\centering
\includegraphics[width=0.48\textwidth]{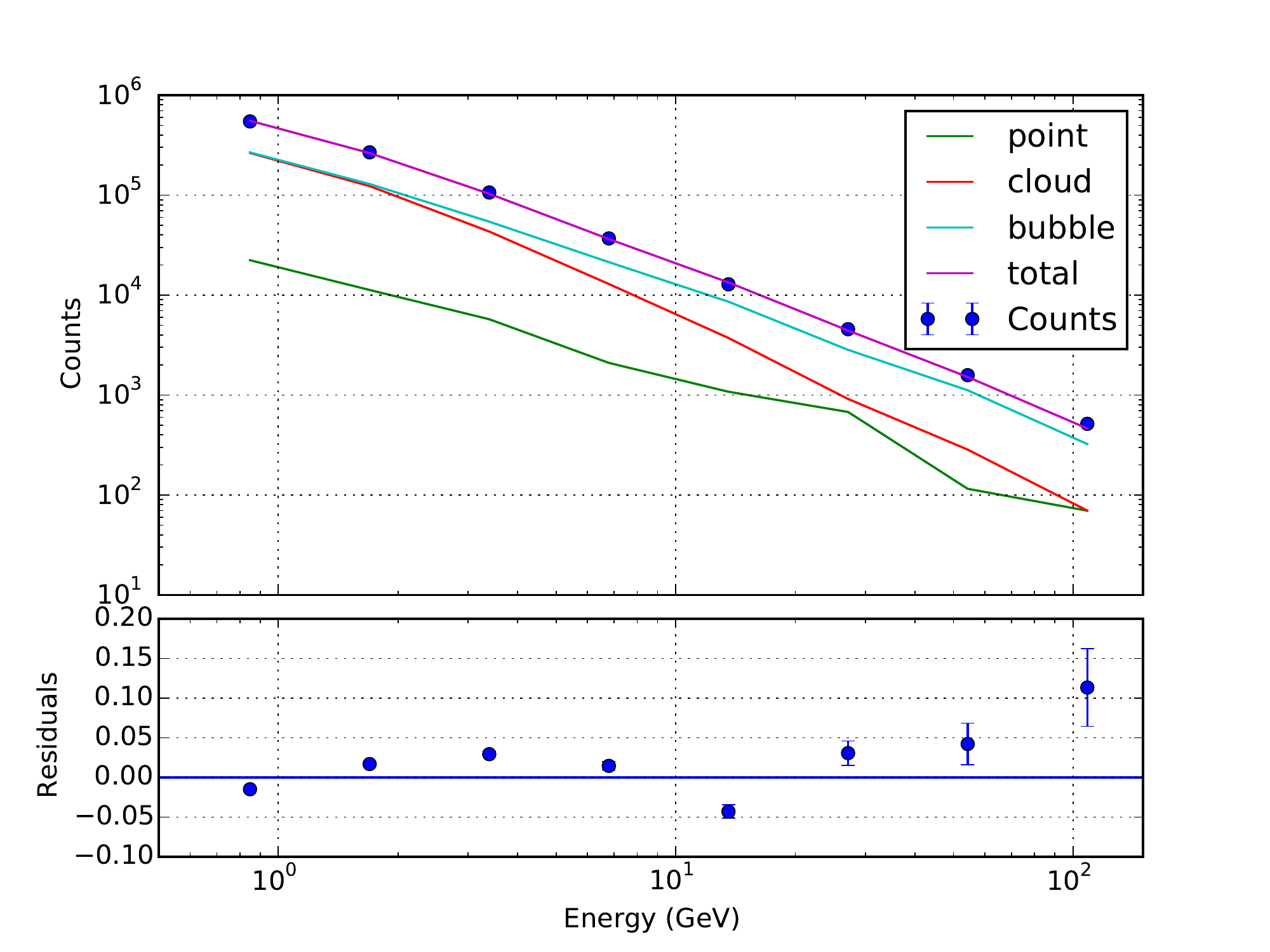}
\includegraphics[width=0.48\textwidth]{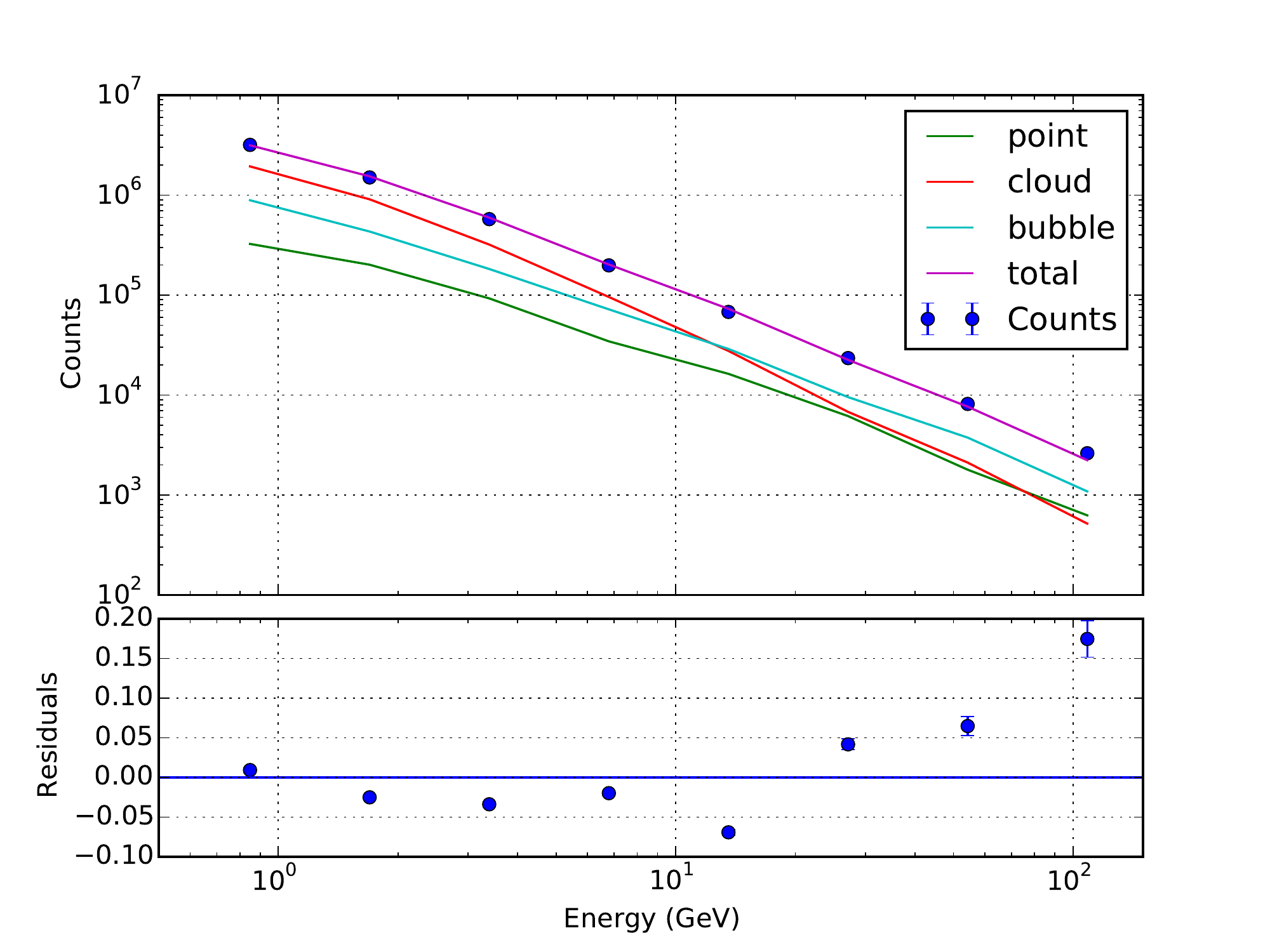}
\caption{Photon counts and relative residuals for a purely astrophysical sky
model within our primary ROI (left) and test ROI (right). 
Here, cloud and bubble refer to the emission components with cloud-like and bubble-like spectra identified in \cite{Selig:2014qqa}, respectively.}
\label{fig:counts}
\end{figure*}

First, we investigate the possibility for a purely astrophysical $\gamma$-ray
sky without DM annihilation signatures, which means to set $n_{dm}^{ijk}=0$
while fitting the remaining model parameters $\alpha$ and $\beta$
for all locations $i\in\mathrm{ROI}$. Fig.~\ref{fig:counts} shows
the observed and modeled counts within the two ROIs described above
as well as the residuals between model and data. In both ROIs, a purely
astrophysical DGE model fits the data reasonably well. The largest
residual appears in the highest used energy bin. There, the limited
photon statistics might still cause problems to D$^{3}$PO in separating
point sources from diffuse emission as it clearly has done at the
next higher energy bin. Therefore, we do not consider the discrepancy
between model and data at this energy as an serious indicator of DM
or other new physics. However, a small, but significant photon count
excess around several GeV can be recognized as well for our primary
ROI, but not for the test ROI. This excess seems to be the GeV excess
reported in the literature and might be a possible DM annihilation
signature.

\subsection{DM annihilation }
\begin{figure*}[!htb]
\centering
\includegraphics[width=0.48\textwidth]{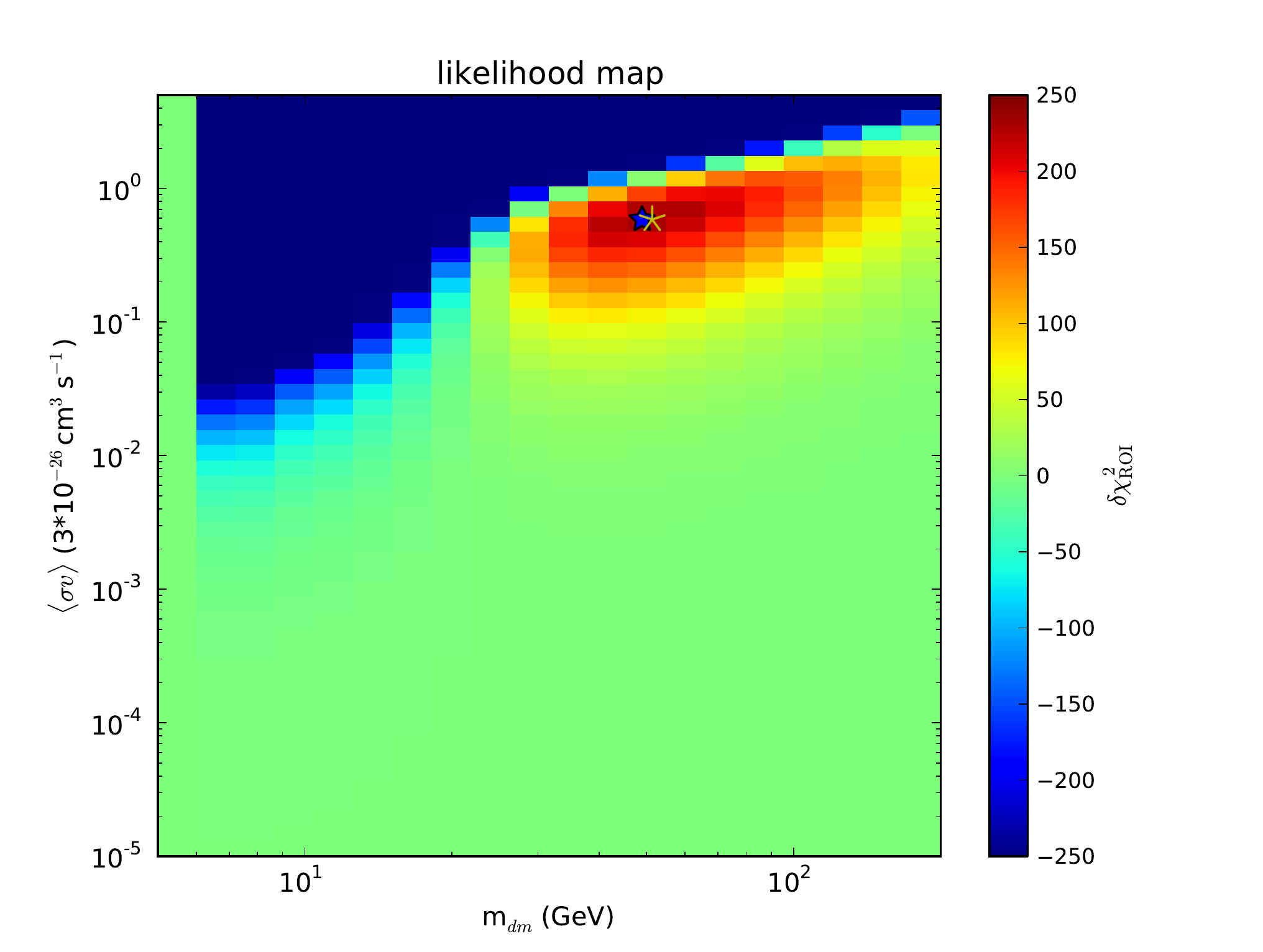}
\includegraphics[width=0.48\textwidth]{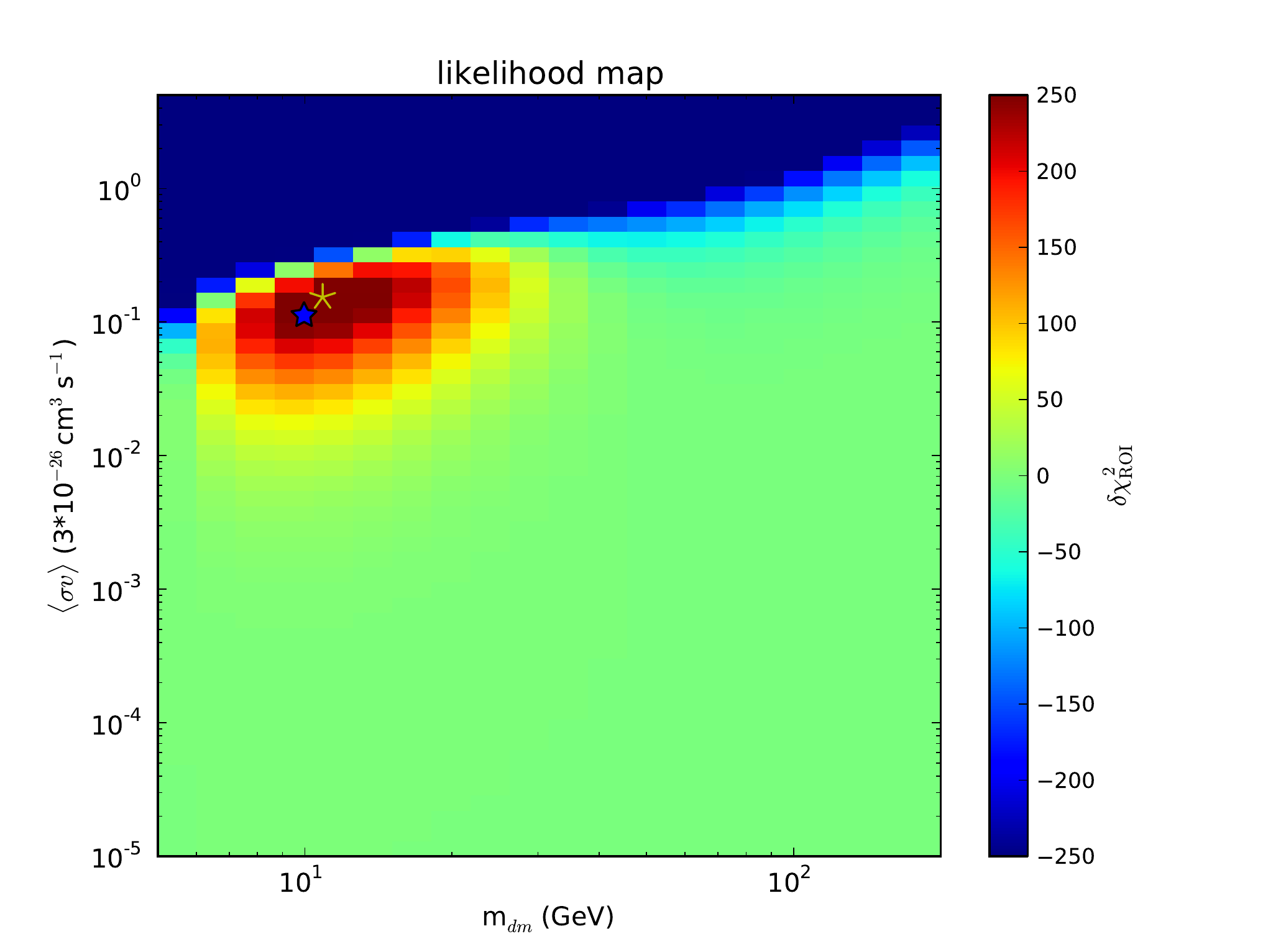}
\caption{The likelihood for $b\bar{b}$ (left) and for $\tau^{-}\tau^{+}$
(right) final annihilation states in terms of $\delta\chi_{\mathrm{ROI}}^{2}(m_{\mathrm{dm}},\langle\sigma v\rangle)$
given by Eq. \ref{eq:ts} for the primary ROI. The blue stars 
indicate best fit values from \citep{Calore:2014xka} and the yellow
stars are our best fit DM parameters $(m_{\mathrm{dm},}\langle\sigma v\rangle)_{\star}$.}
\label{fig:likelihood map}
\end{figure*}

\begin{figure*}[!htb]
\centering
\includegraphics[width=0.48\textwidth]{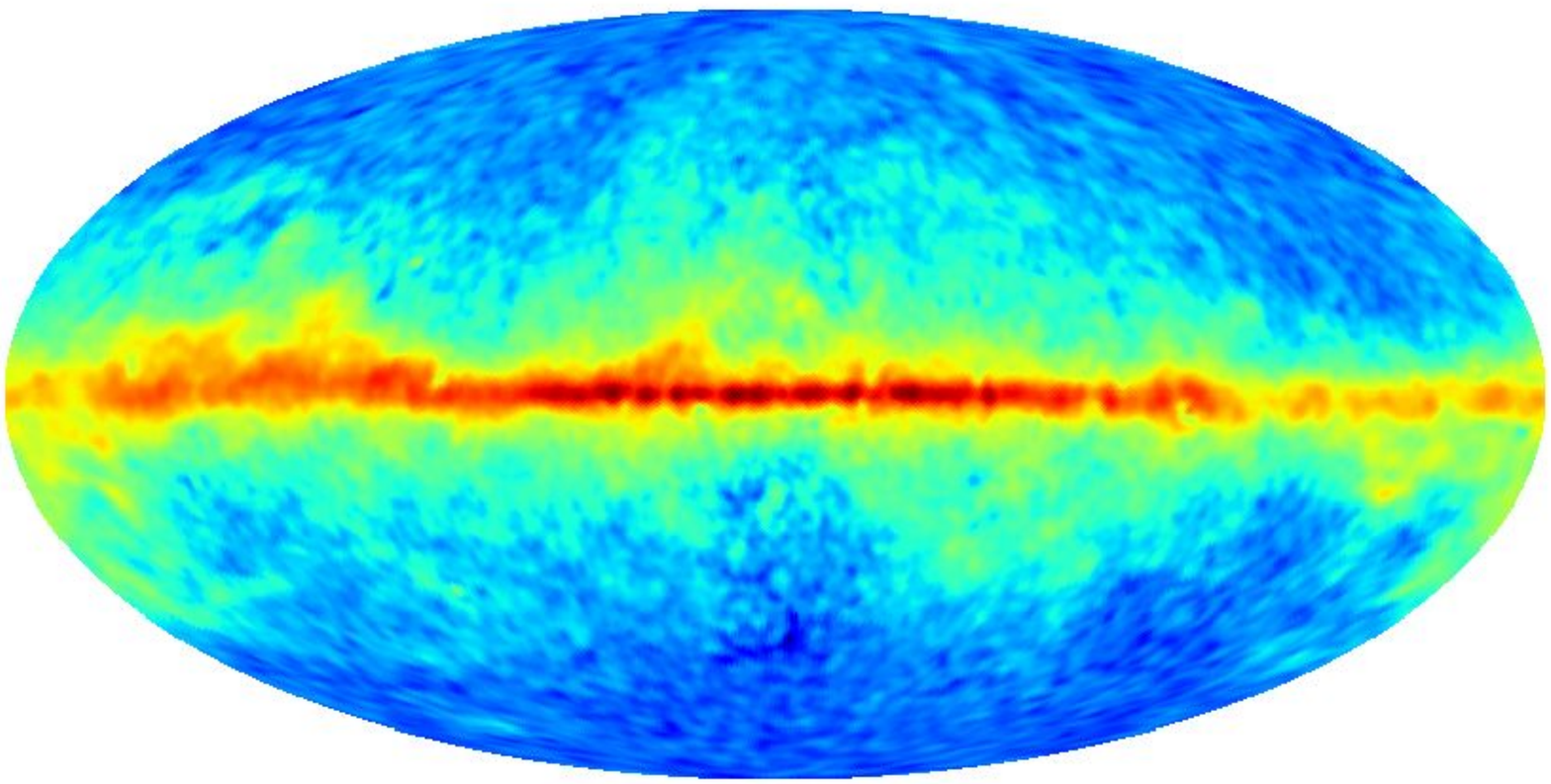}
\includegraphics[width=0.48\textwidth]{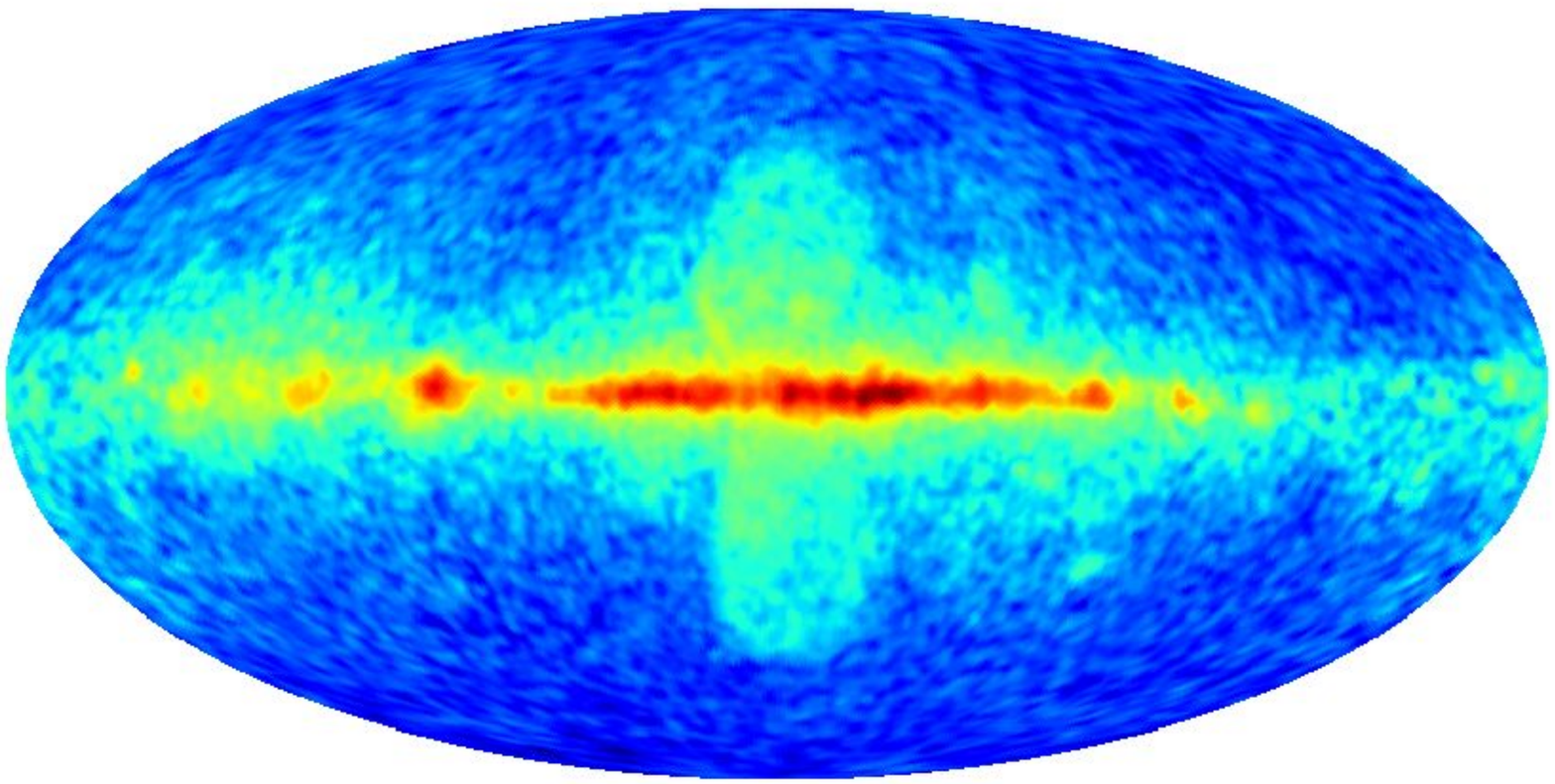}
\caption{Cloud-like (left) and bubble-like (right) gamma-ray emission components
in logarithmic units after permitting for the presence of DM annihilation
for the best fit DM model with the $b\bar{b}$ annihilation channel.}
\label{fig:Cloud-like-(left)-and}
\end{figure*}

To investigate whether the GeV photon count excess observed within
our primary ROI could be caused by DM annihilation, we scan dark the
matter parameters $m_{\mathrm{dm}}$ and $\langle\sigma v\rangle$
while fitting the astrophysical DGE parameter sets $\alpha$ and $\beta$.
The corresponding improvement of our objective function 
\begin{eqnarray}
 &  & \delta\chi_{\mathrm{ROI}}^{2}(m_{\mathrm{dm}},\langle\sigma v\rangle)=\nonumber \\
 &  & \mathrm{min}_{\alpha,\beta}\chi_{\mathrm{ROI}}^{2}(0,0,\alpha,\beta)-\mathrm{min}_{\alpha,\beta}\chi_{\mathrm{ROI}}^{2}(m_{\mathrm{dm}},\langle\sigma v\rangle,\alpha,\beta)\label{eq:ts}
\end{eqnarray}
thanks to the presence of DM is shown in Fig.~\ref{fig:likelihood map}
for the $b\bar{b}$ and $\tau^{-}\tau^{+}$ final state DM annihilation
chains. Within this we also identify the best fit DM parameters $(m_{\mathrm{dm},}\langle\sigma v\rangle)_{\star}$
with maximum $\delta\chi_{ROI}^{2}$. These best fit locations agree
well with the ones found by Calore et al.  \cite{Calore:2014xka}, who consider
DGE model uncertainties and also empirical model systematics in their
analysis. As a sanity check, one can verify that the resulting cloud-like
and bubble-like components (Fig.~\ref{fig:Cloud-like-(left)-and})
do not seem obviously be disturbed by the presence of DM annihilation.
For instance, there is no evident emission decrement at the GC apparent on these maps.
Had there been one, it would have been an indication of the DM component fitting away 
the astrophysical ones.

\begin{figure*}[!htb]
\centering
\includegraphics[width=0.48\textwidth]{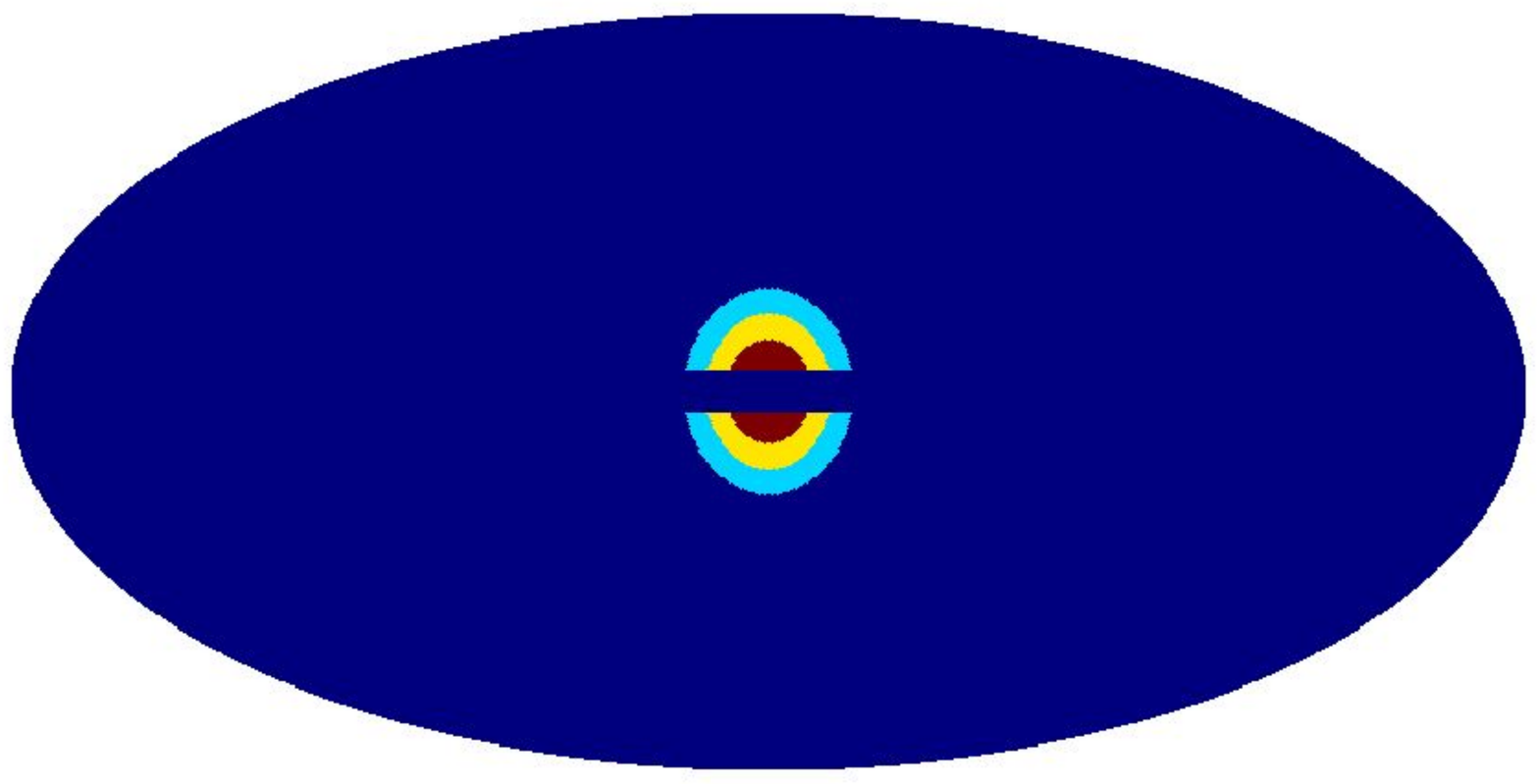}
\includegraphics[width=0.48\textwidth]{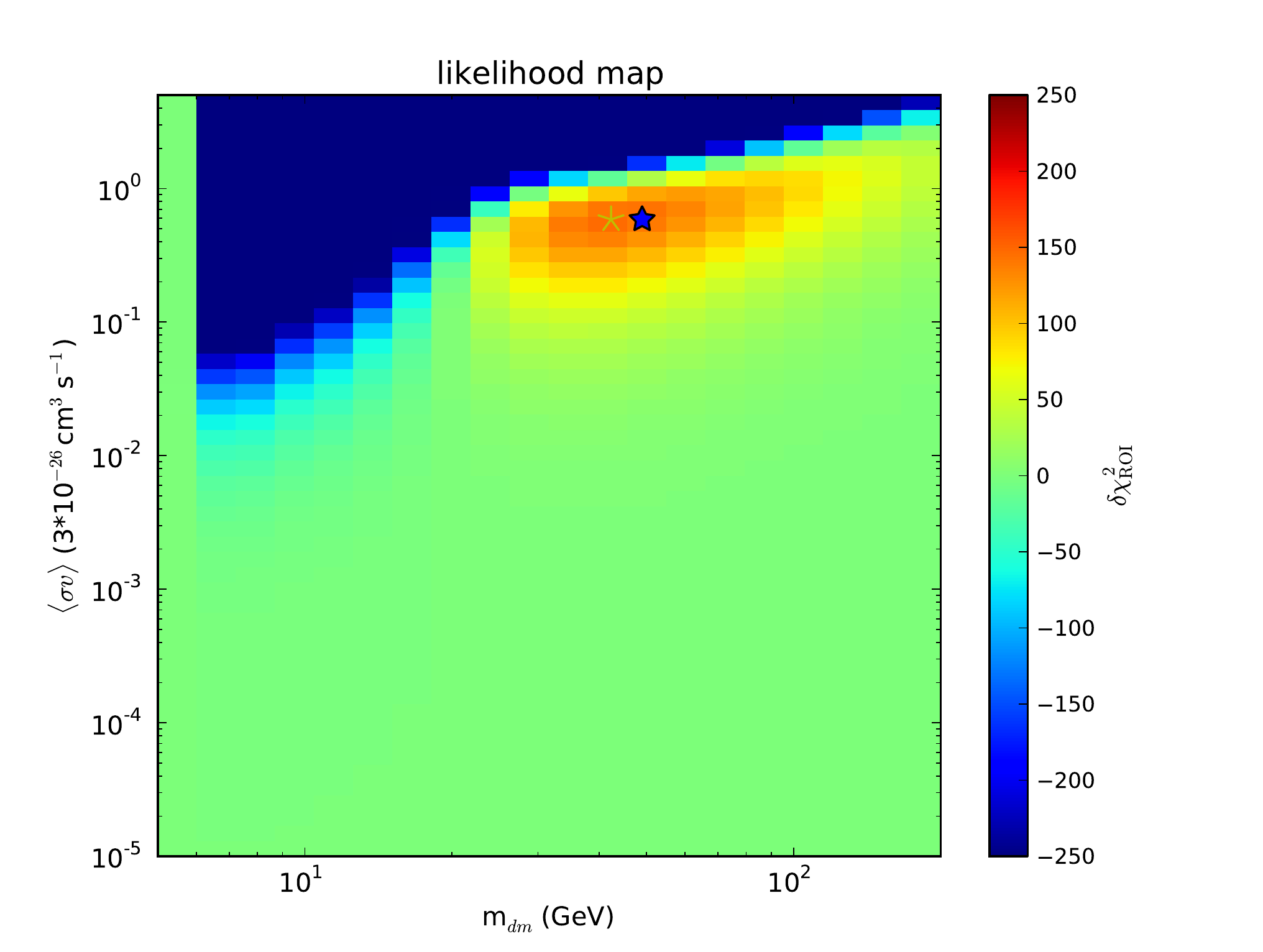}\\
\includegraphics[width=0.48\textwidth]{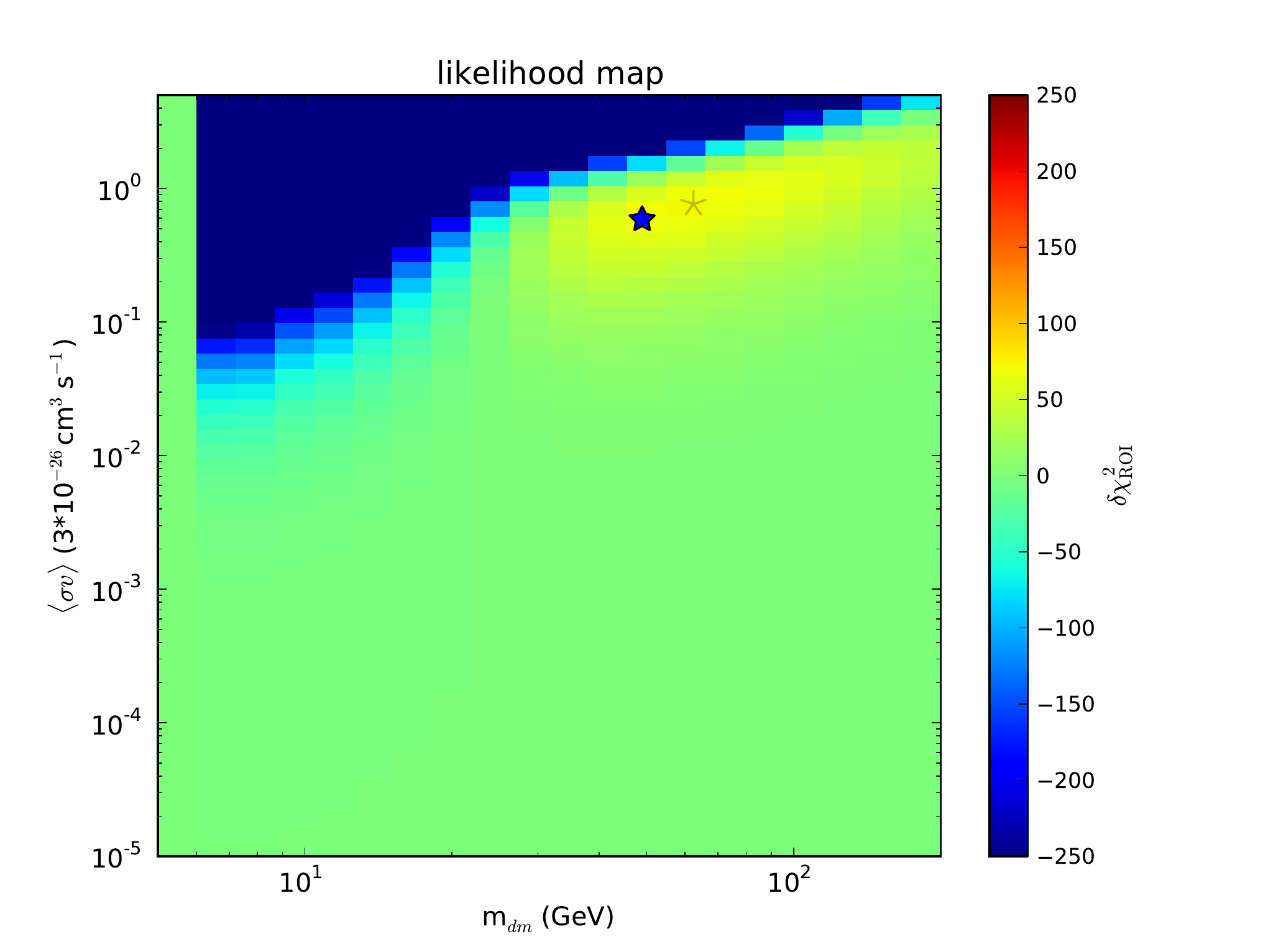}
\includegraphics[width=0.48\textwidth]{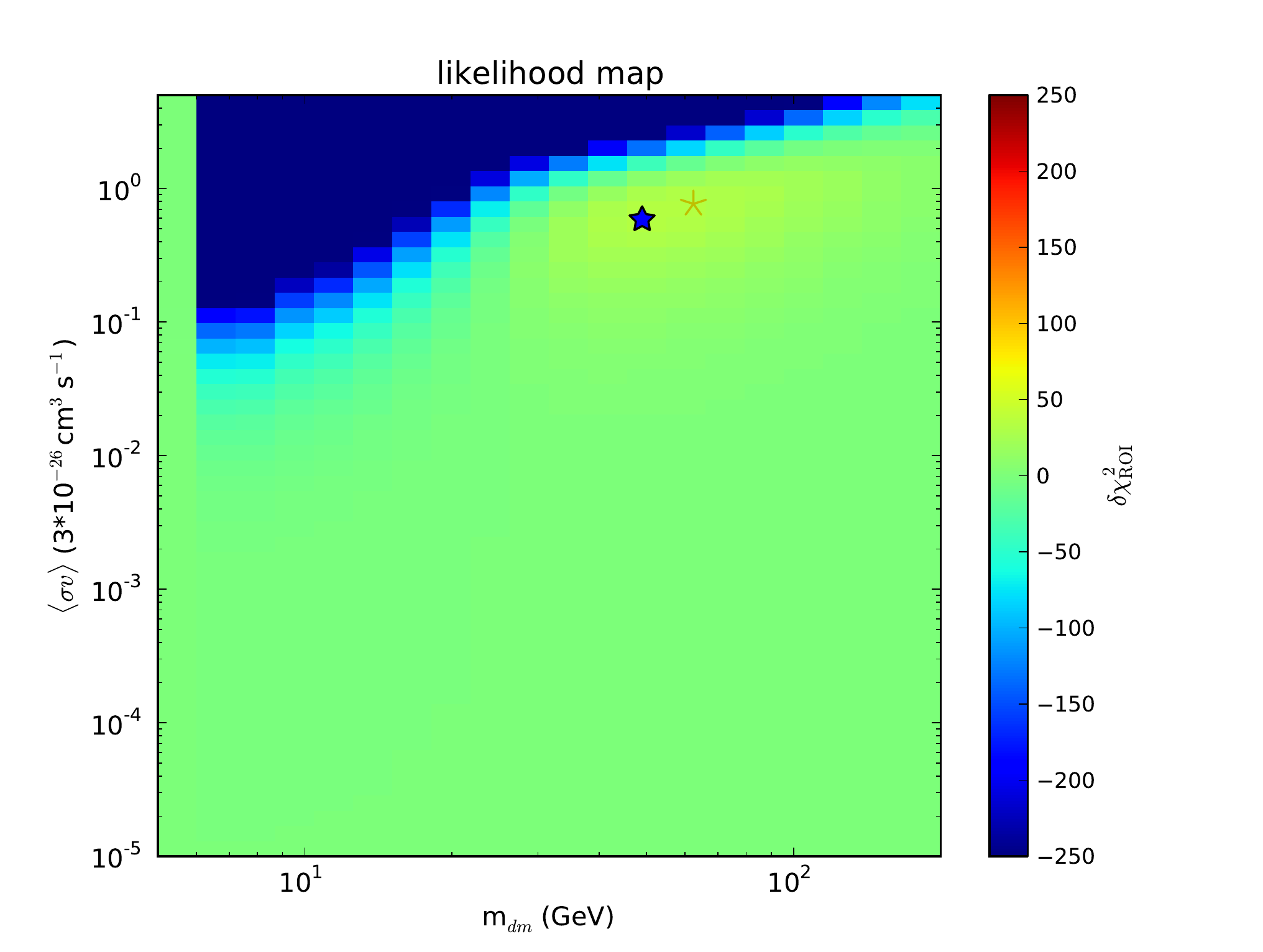}
\caption{Top left: Different ROIs to test the consistency of the best fit DM
parameters $(m_{\mathrm{dm},}\langle\sigma v\rangle)_{\star}$. Top
right to bottom right: Corresponding $\delta\chi_{\mathrm{ROI}}^{2}(m_{\mathrm{dm}},\langle\sigma v\rangle)$
given by Eq. \ref{eq:ts} for $b\bar{b}$ final annihilation states
for these ROIs from inside out. Here the blue stars 
indicate best fit values from \citep{Calore:2014xka} and the yellow
stars are our best fit DM parameters for the corresponding ROIs.}
\label{fig:different_regions}
\end{figure*}

As a further check, we change the ROIs to verify that the best fit
DM parameters $(m_{\mathrm{dm},}\langle\sigma v\rangle)_{\star}$
inferred from different regions are consistent with each other. We
choose three regions with different angular distances to the GC: from
0 to 10 degrees, 10 to 15 degrees and 15 to 20 degrees, and in each region we also mask out the inner 4 degrees of the galactic disk in latitude, see Fig.~\ref{fig:different_regions}.
The corresponding likelihood maps for $b\bar{b}$ annihilation final
states (also Fig.~\ref{fig:different_regions}) are consistent with
each other and exhibit only slightly changing best fit DM parameters
$(m_{\mathrm{dm},}\langle\sigma v\rangle)_{\star}$.

In order to test for a potential astrophysical, non-DM annihilation
related origin of this signal we try to identify the sky locations
driving $\delta\chi_{\mathrm{ROI}}^{2}$ . To this end we construct
all sky maps of $\delta\chi_{\star}^{2}$ for the best fit DM parameters
$(m_{\mathrm{dm}\star,}\langle\sigma v\rangle_{\star})$ as 
\begin{equation}
\delta\chi_{\star i}^{2}=\mathrm{min}_{\alpha_{i},\beta_{i}}\chi_{i}^{2}(0,0,\alpha_{i},\beta_{i})-\mathrm{min}_{\alpha_{i},\beta_{i}}\chi_{i}^{2}(m_{\mathrm{dm}\star},\langle\sigma v\rangle_{\star},\alpha_{i},\beta_{i}).\label{eq:tsi}
\end{equation}

\begin{figure*}[!htb]
\centering
\includegraphics[width=0.48\textwidth]{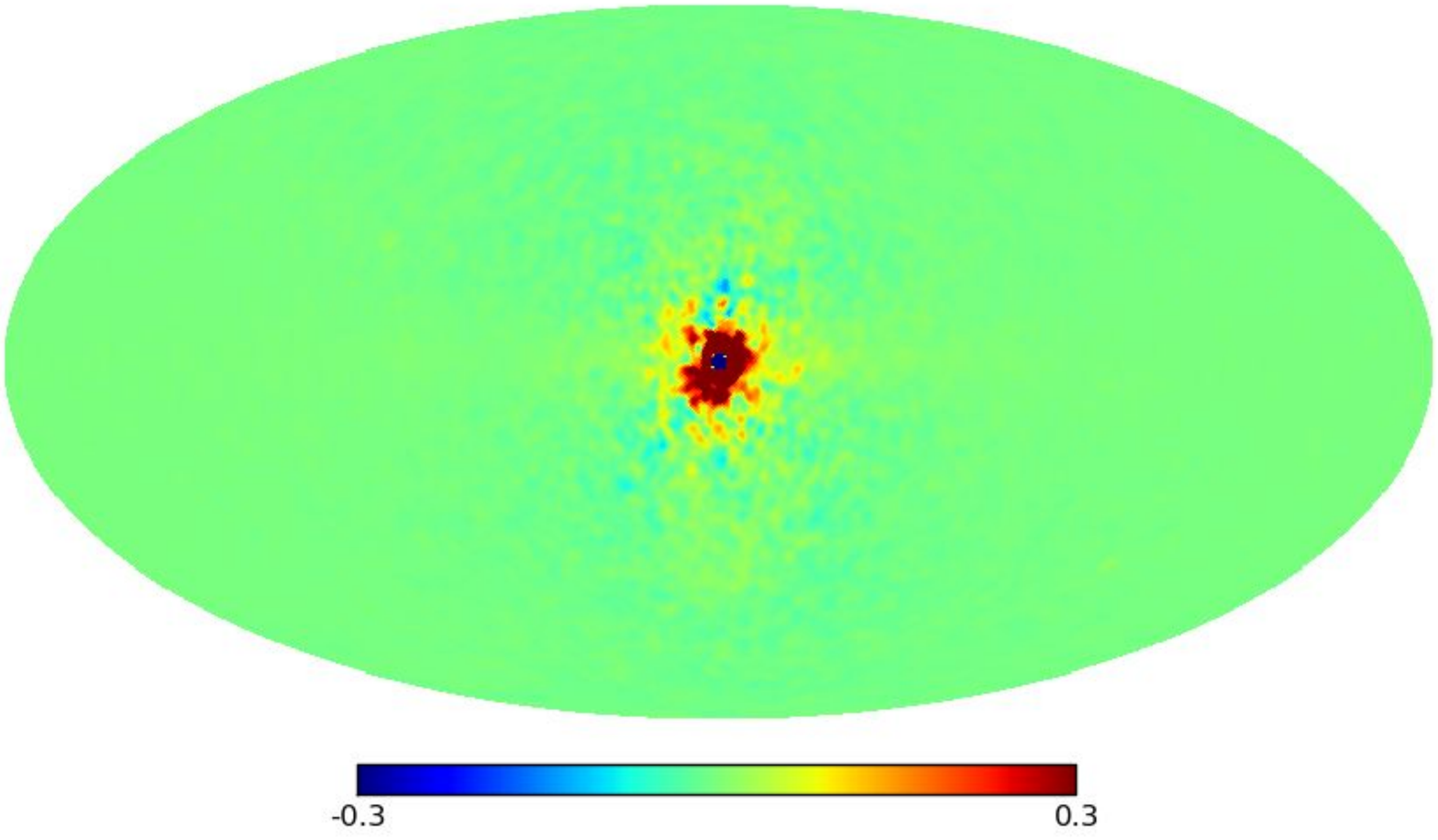}
\includegraphics[width=0.48\textwidth]{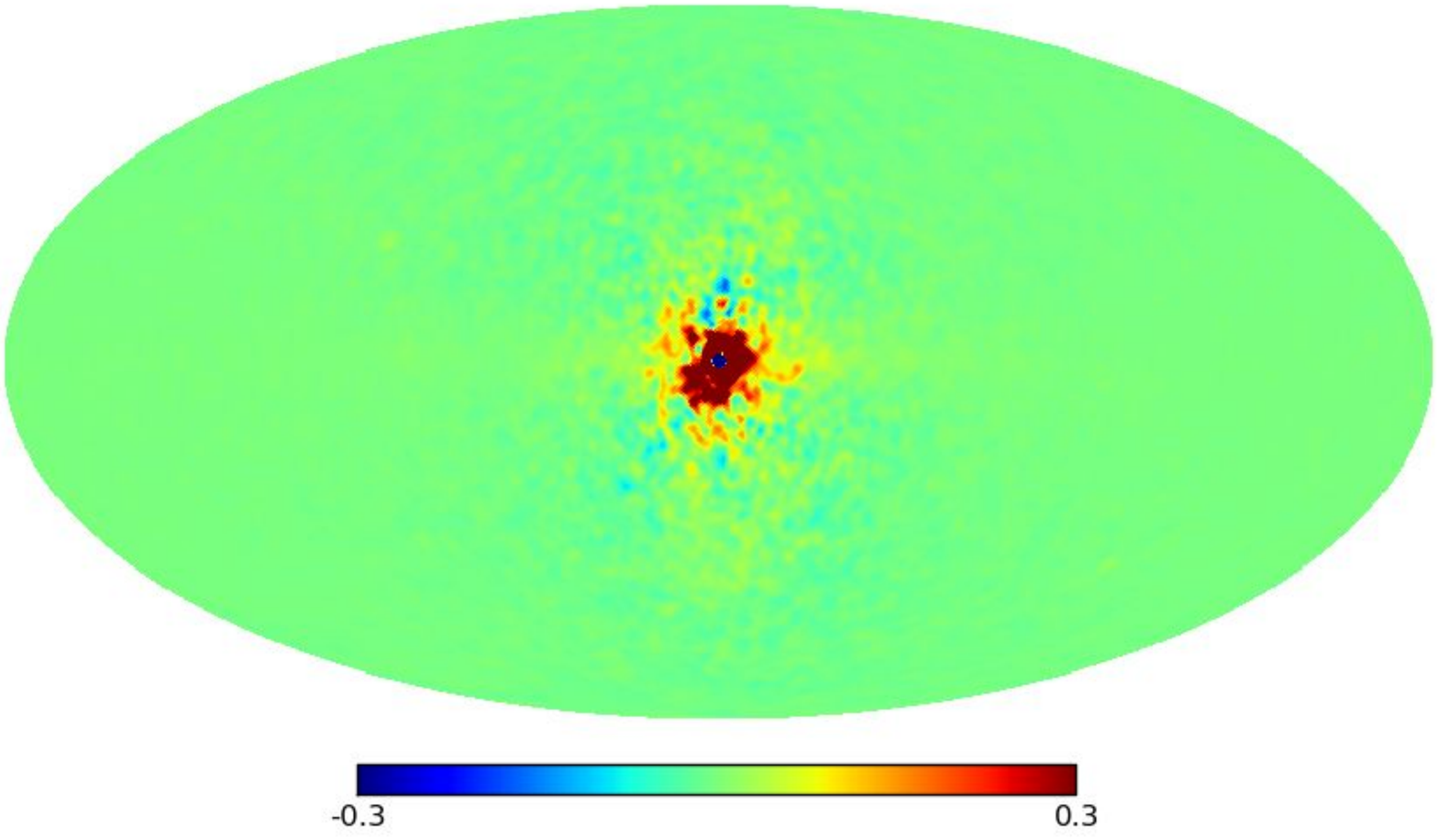}
\caption{The map of the likelihood improvements $\delta\chi_{\star i}^{2}$
due to a DM components with properties $(m_{\mathrm{dm},}\langle\sigma v\rangle)_{\star}$
according to Eq. \ref{eq:tsi} for $b\bar{b}$ (left) and $\tau^{-}\tau^{+}$
(right) final annihilation states.}
\label{fig:pixel_by_pixel_ts}
\end{figure*}

In case such a map exhibits the morphology of a known galactic structure,
like the Fermi bubbles, the molecular clouds, or others, this structure
will be suspected to be the origin of an only apparent DM signal.
Fig.~\ref{fig:pixel_by_pixel_ts} shows that the improvement in $\chi^{2}$
due to the inclusion of DM annihilation $\gamma$-rays is almost spherically
distributed around the GC. This $\chi^{2}$ improvement is also seen
at the regions around GC, which were excluded from the fit. This is
as it should be in case this radiation was a missing element of the
$\gamma$-ray sky. The pixels at the very center of the Galaxy do
not request a DM component. However, the strong CR accelerators there
with different spectra compared to our simplistic DGE components might
have confused the fit.

\begin{figure*}[!htb]
\centering
\includegraphics[width=0.48\textwidth]{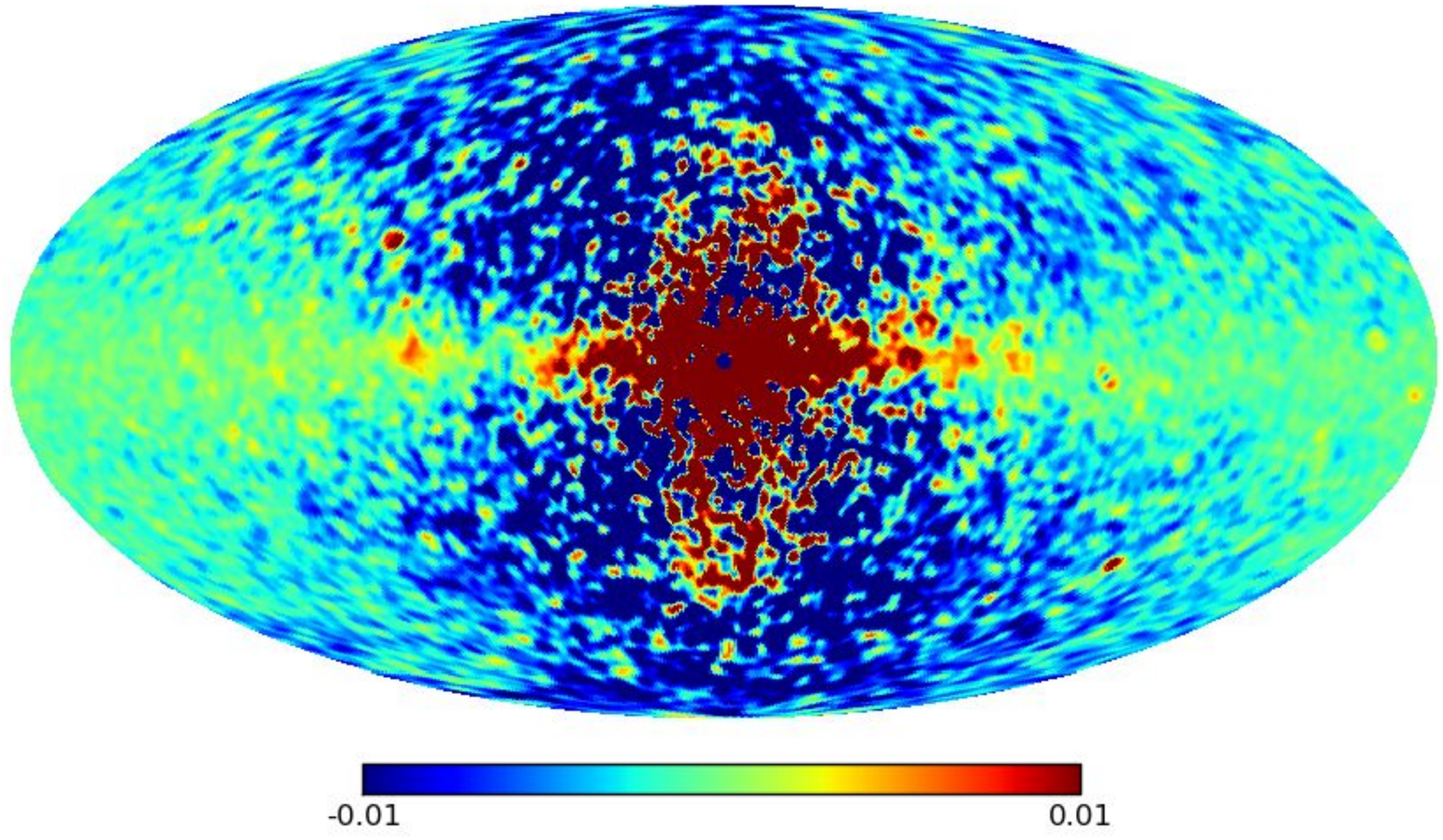}
\caption{Like left panel of Fig.~\ref{fig:pixel_by_pixel_ts}, but with rescaled
and strongly saturated color scale to highlight non-central regions.}
\label{fig:pixel_by_pixel_ts_rescale}
\end{figure*}

Anyhow, an inspection of the more subtle $\chi^{2}$-changes made
visible by tuning the colorbar (Fig.~\ref{fig:pixel_by_pixel_ts_rescale})
reveals the morphology of the Fermi bubbles as well as of the galactic
disk in $\delta\chi_{\star i}^{2}$ structures at locations more distant
from the GC. This suggests a single ``bubble-like'' spectra to be
an imperfect representation of the hot ISM $\gamma$-emission spectrum
and that existing variations therein have spectral similarity with
the spectrum of DM annihilation. The contribution of these morphologically
suspect regions to the total $\delta\chi_{\mathrm{ROI}}^{2}$ is marginal,
but could indicate a problem also prevailing within our primary ROI.

\begin{figure*}[!htb]
\centering
\includegraphics[width=0.48\textwidth]{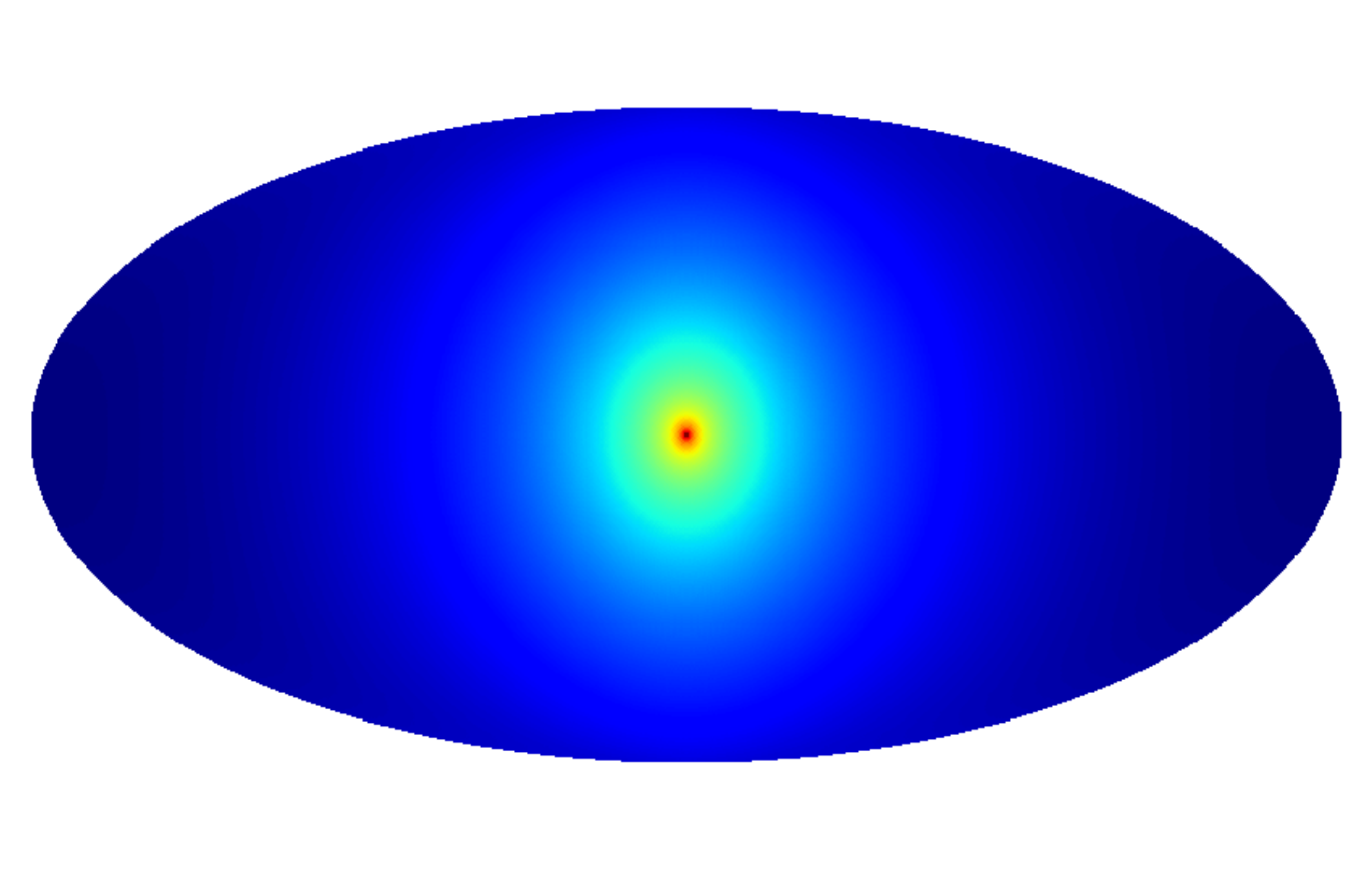}
\includegraphics[width=0.48\textwidth]{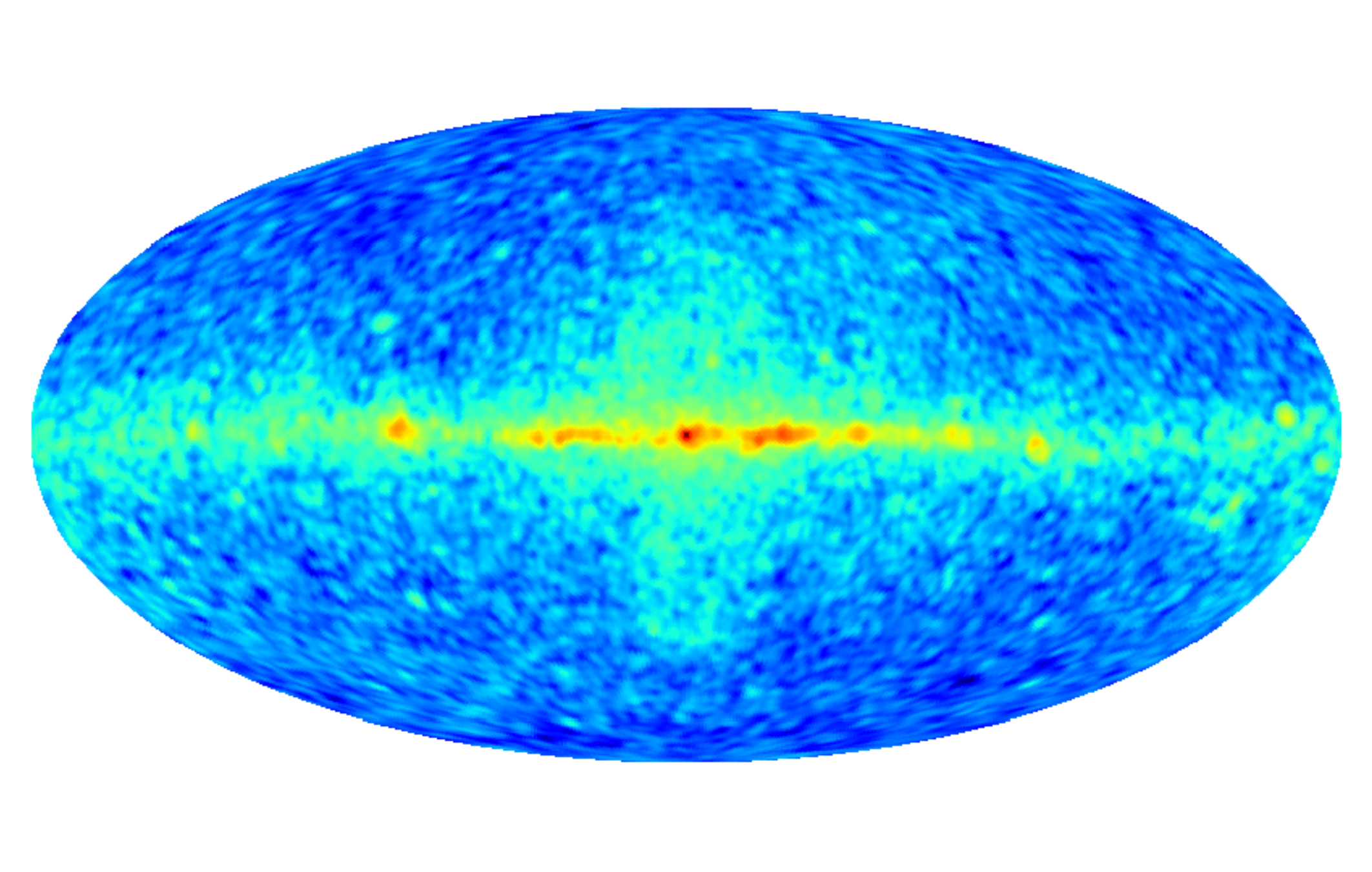}
\caption{In the left panel we show the anticipated spatial distribution of
the number of counts $n_{\mathrm{DM}}^{i}$ from DM annihilation,
and in the right panel we show $\gamma_{i}n_{\mathrm{DM}}^{i}$, which
indicate the spatial distribution of number of counts from fitting
with new parameter $\gamma_{i}$.}
\label{fig:norm}
\end{figure*}

A spectral component resembling DM annihilation improves the fit also
at locations where not much DM annihilation is expected. In order
to investigate this a bit further, the following experiment is performed:
The spatial DM template is allowed change by introducing modification
factors $\gamma_{i}$ via the replacement $n_{\mathrm{dm}}^{ijk}\rightarrow\gamma_{i}n_{\mathrm{dm}}^{ijk}$
in Eq.~\ref{eq:lambdda}. Fig.~\ref{fig:norm} compares the original
$n_{\mathrm{dm}}^{i}$-map with the $\gamma_{i}n_{\mathrm{dm}}^{i}$-map
resulting from a simultaneous fit in $\alpha$, $\beta$, and $\gamma$
with the DM parameters are fixed to $(m_{\mathrm{dm},}\langle\sigma v\rangle)_{\star}$
and assuming $b\bar{b}$ annihilation final states. A numerical comparison
of the Galactic inner regions of these maps shows that the spatial
DM template is relatively good, as the modification factors have values
around unity in there, which is consistent with the result in Fig.~
\ref{fig:likelihood map}. 

However, Fig.~\ref{fig:norm} also shows that a spectral component
mimicking DM annihilation is strongly wanted by the data for regions
without much expected DM signal. In particular, regions hosting the
hot interstellar medium seem to request such a DM-like spectral component
(compare to Fig.~\ref{fig:Cloud-like-(left)-and}).  This kind of possible bipolar structures is also favored in recent work \cite{Yang:2016duy,Carlson:2016iis} for the morphology of GeV excess. This is, on
the one hand, an indicator that our simplistic two component galactic
model is not complex enough to fully represent the inverse Compton
emission of the Galaxy. Given the short cooling times of the relativistic
electrons involved in this channel, some spectral variations in their
emission is not too surprising. On the other hand, the preferred dilute
ISM-like morphology of a DM-like spectral component disfavors the
DM annihilation scenario as well as the possibility of a blend of
weak $\gamma$-ray emitting point sources for their expected very
different $\gamma$-ray morphologies.

Previously proposed astrophysical explanations, such as extra CR in
the GC producing leptons \citep{Petrovic:2014uda,Cholis:2015dea,Gaggero:2015nsa}
could possibly contribute to the bubble shape structures, and a population
of Galactic millisecond pulsars \citep{Abazajian:2012pn,Gordon:2013vta,Mirabal:2013rba,Yuan:2014rca,Calore:2014oga,2015arXiv150605124L,2015arXiv150605104B,Hooper:2013nhl,Cholis:2014lta}
could in principle also contribute to the $\gamma$-ray emission Galactic
disk.

\subsection{Upper limits}

\begin{figure*}[!htb]
\centering
\includegraphics[width=0.48\textwidth]{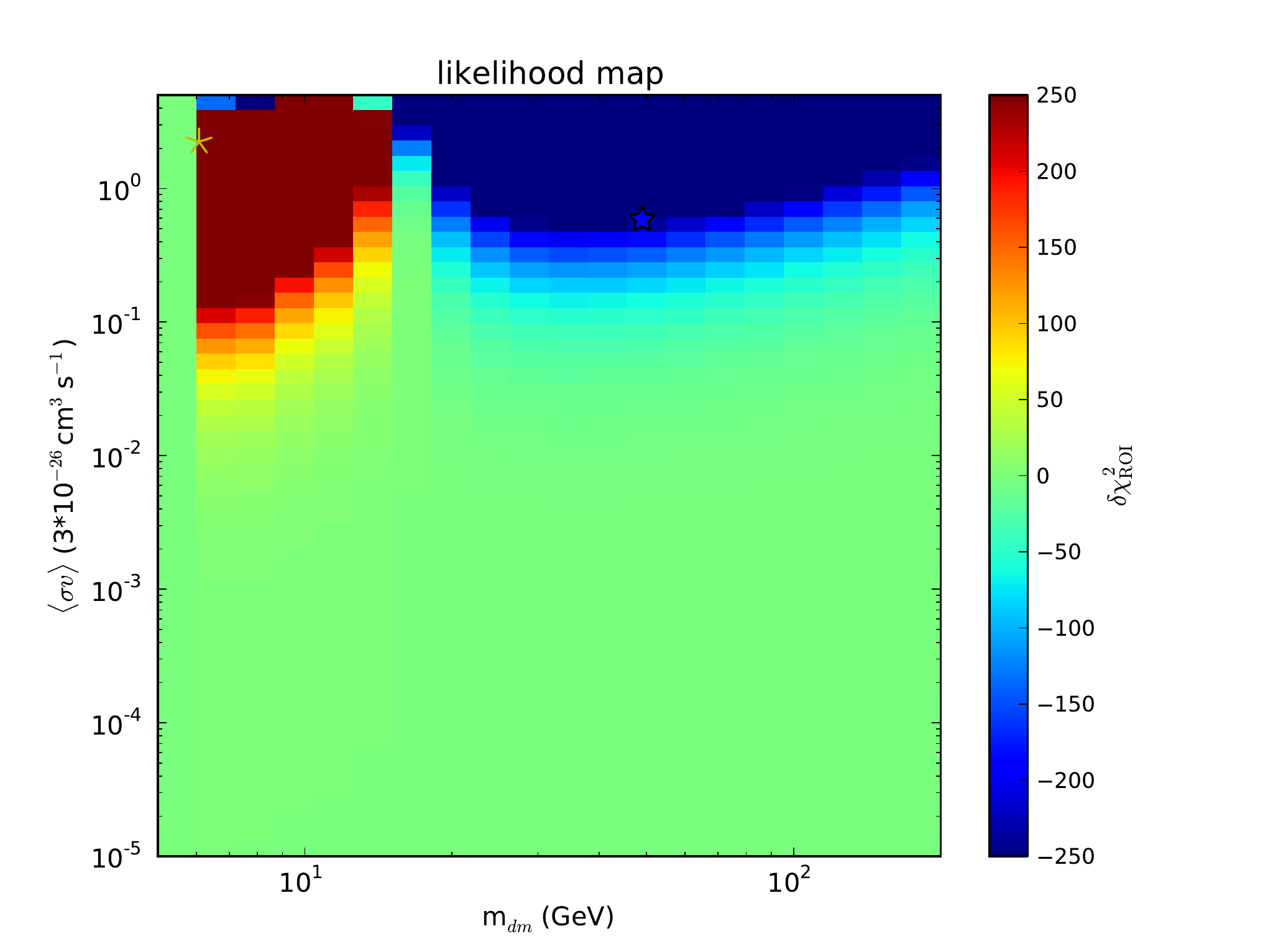}
\includegraphics[width=0.48\textwidth]{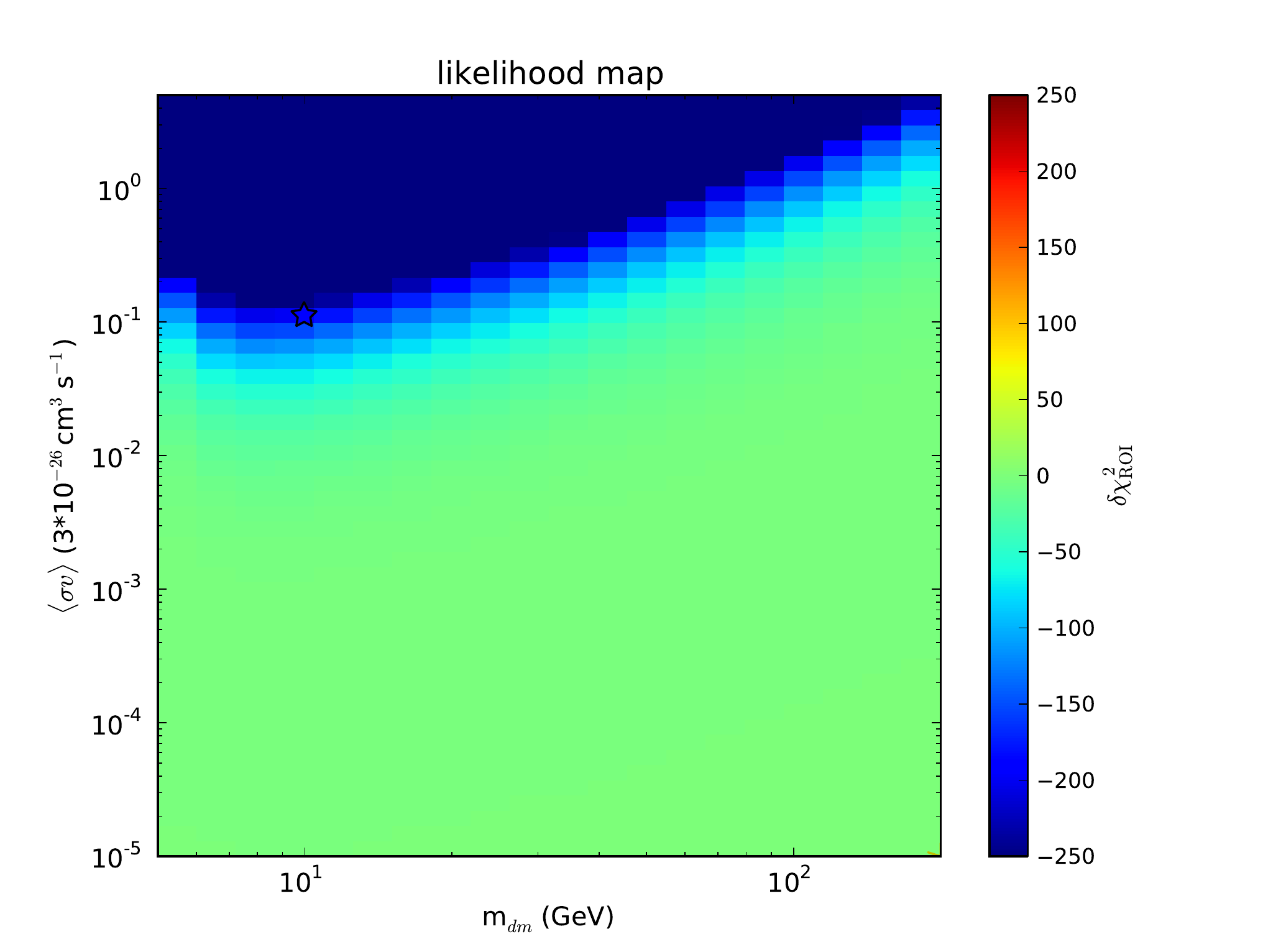}
\caption{The likelihood for $b\bar{b}$ (left) and for $\tau^{-}\tau^{+}$
(right) annihilation final states in terms of $\delta\chi_{\mathrm{ROI}}^{2}(m_{\mathrm{dm}},\langle\sigma v\rangle)$
given by Eq. \ref{eq:ts} for the test ROI. The blue stars indicate
best fit values from \citep{Calore:2014xka} and the yellow stars
are for the test ROI our best fit DM parameters $(m_{\mathrm{dm},}\langle\sigma v\rangle)_{\star}$.}
\label{fig:anticenter}
\end{figure*}

Large DM annihilation cross sections can be excluded, as they would
imply $\gamma$-ray counts strongly in excess to the data. Here, we
switch back to assuming the DM to have spatially a generalized NFW
profile. Upper limits on $\langle\sigma v\rangle$ should be placed
(as a function of DM mass $m_{\mathrm{dm}}$ and final state channel
of the annihilation) at the transition from green to blue in Fig.~
\ref{fig:likelihood map} and Fig.~\ref{fig:anticenter} for the primary
and test ROI, respectively.\footnote{Specifying a formal statistical upper limit curve would pretend a
not justified precision since this work shows that the DGE model systematics
dominate the uncertainty budget.} 

It is interesting to note that the test ROI seems to be far more exclusive
for DM models compared to previous work deriving upper limits from high latitude region \cite{Calore:2013yia, Cholis:2013ena,DiMauro:2015tfa,Fornasa:2015qua}, if we ignore the DM-affine region in the parameter
space for $b\bar{b}$ visible in Fig.~\ref{fig:anticenter}, which
is driven by an excess in the lowest energy bin only (see Fig.~\ref{fig:counts}).
These exclusion limits are so thigh that they seem to exclude the
best fit scenarios for our primary ROI. These limits, however, depend
strongly on the validity of the adopted NFW profile in the outer Galaxy,
which certainly can be questioned. 

\begin{figure*}[!htb]
\centering
\includegraphics[width=0.48\textwidth]{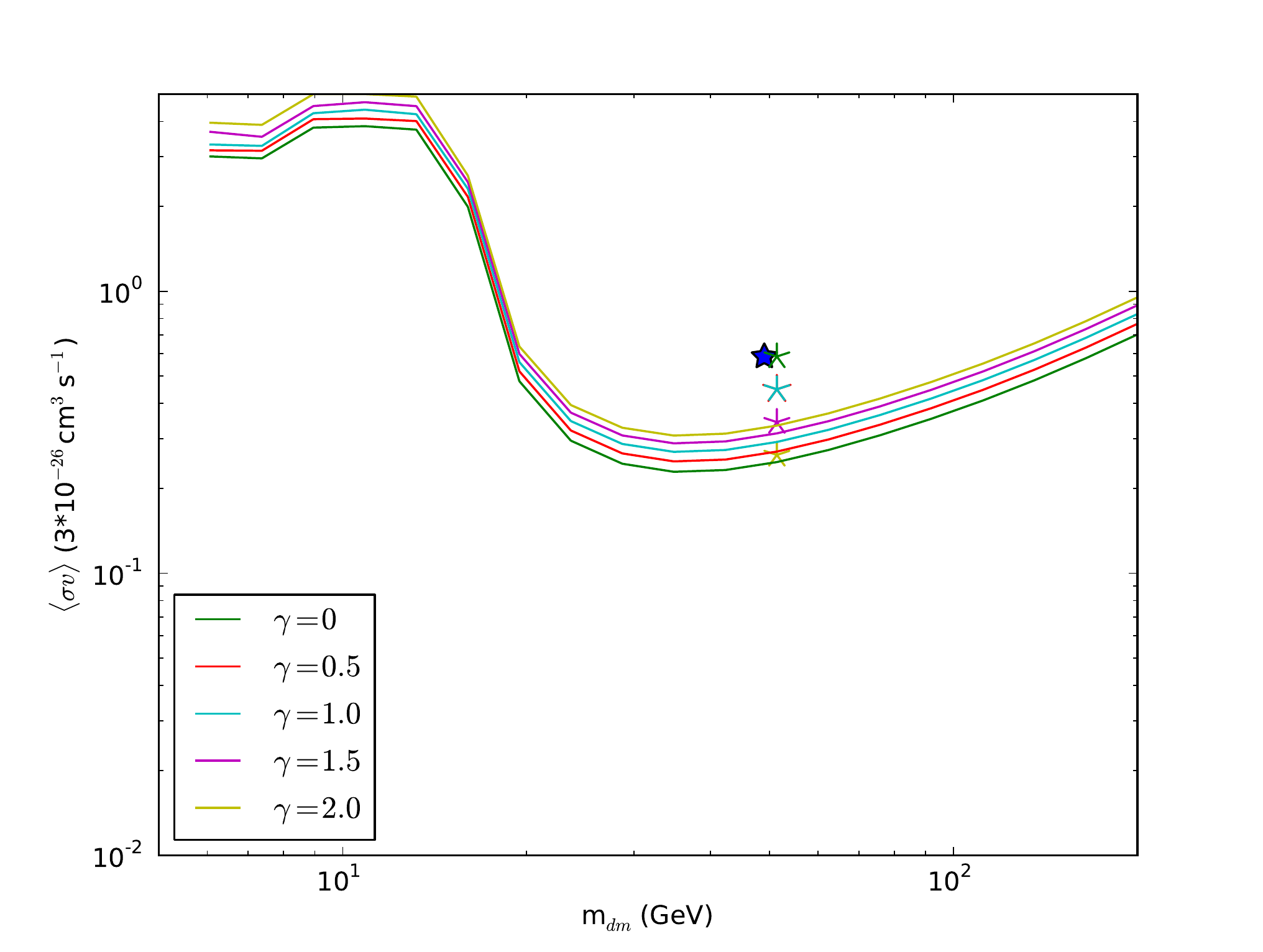}
\includegraphics[width=0.48\textwidth]{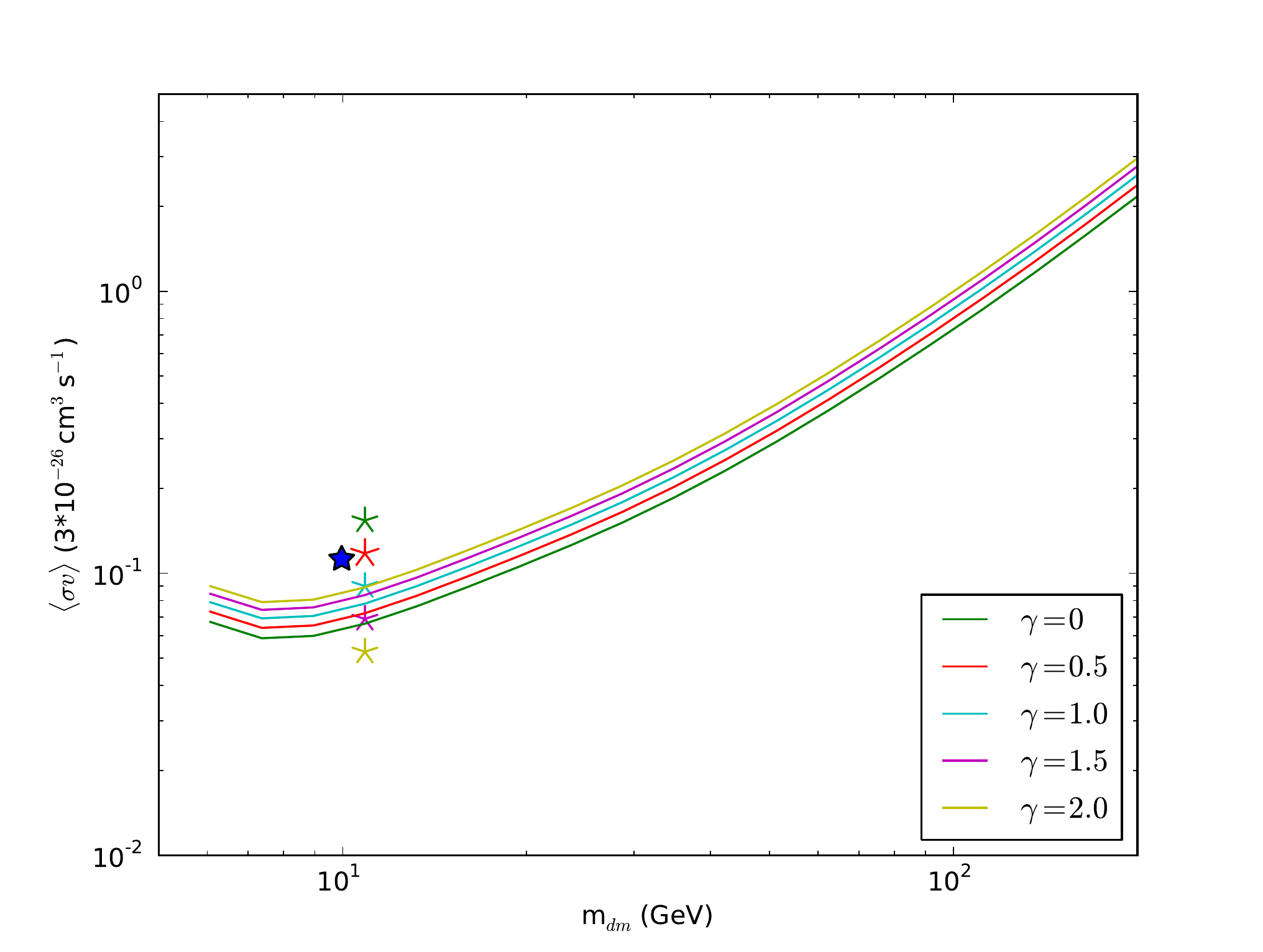}
\caption{
Impact of the assumed Galactic DM profile for different values of $\gamma$ coded in different colors. The best fit DM parameters derived within the primary ROI (symbols) and upper limits from test ROI (lines) for $b\bar{b}$ (left) and for $\tau^{-}\tau^{+}$ (right) annihilation final states are shown. 
The blue stars are best fit values from \citep{Calore:2014xka}. Colored lines are upper limits derived from the test ROI at $\Delta \chi^2=-100$. The best fit parameters from the primary ROI are marked as stars.}
\label{fig:gnfw_limit}
\end{figure*}

To investigate the robustness of the exclusion limit on the assumed DM profile we generalize the NFW profile to
\begin{equation}
\rho(r)=\frac{\rho_{s}}{(r/r_{s})^{\alpha}(1+r/r_{s})^{3-\alpha+\gamma}}.
\end{equation}
Here, we keep $\alpha=1.2$ to accommodate the GeV excess in the inner Galaxy region, but add the parameter $\gamma$ to describe a possible deviation of the DM profile from NFW in the outer regions of the Galaxy. 
Since we determine the normalization
$\rho_{s}$ by fixing the DM density at the solar radius to $\ensuremath{\rho(r_{\odot}=8.5}\ kpc\ensuremath{)}=0.4\ \mbox{GeV cm}\ensuremath{^{-3}}$ \cite{Catena:2009mf,Salucci:2010qr,Iocco:2011jz} with fixed $r_s=20$ kpc, $\rho_{s}$ changes slightly according to the choice of $\gamma$. 
We repeat our analysis for DM profiles with $\gamma=0.0,$ $0.5$, $1.0$, $1.5$ and $2.0$ in both, the primary and test ROIs. 
The resulting best fit DM parameters of the primary ROI and the upper limits from the test ROI  are shown in Fig.~\ref{fig:gnfw_limit}. 
With increasing $\gamma$ the best fit cross section derived within the primary ROI decreases due to the mentioned change in the DM normalization. Simultaneously, the upper limits derived within the test ROI increases due to the reduced DM density in the outer Galaxy region. 
For the here chosen threshold of $\Delta \chi^2=-100$ a value of  $\gamma \ge 1,5$ is required to make the test ROI upper limits consistent with the primary ROI best fit parameters for both, $b\bar{b}$ and $\tau^-\tau^+$ annihilation final states.

Anyhow, the fact that the galactic
disk desires a DM-like $\gamma$-ray component, whereas the regions
above the disk prefer it to be on a lower level, is again indicative
to an astrophysical nature of this component.

\subsection{Extragalactic component}

\begin{figure*}[!htb]
\centering
\includegraphics[width=0.3\textwidth]{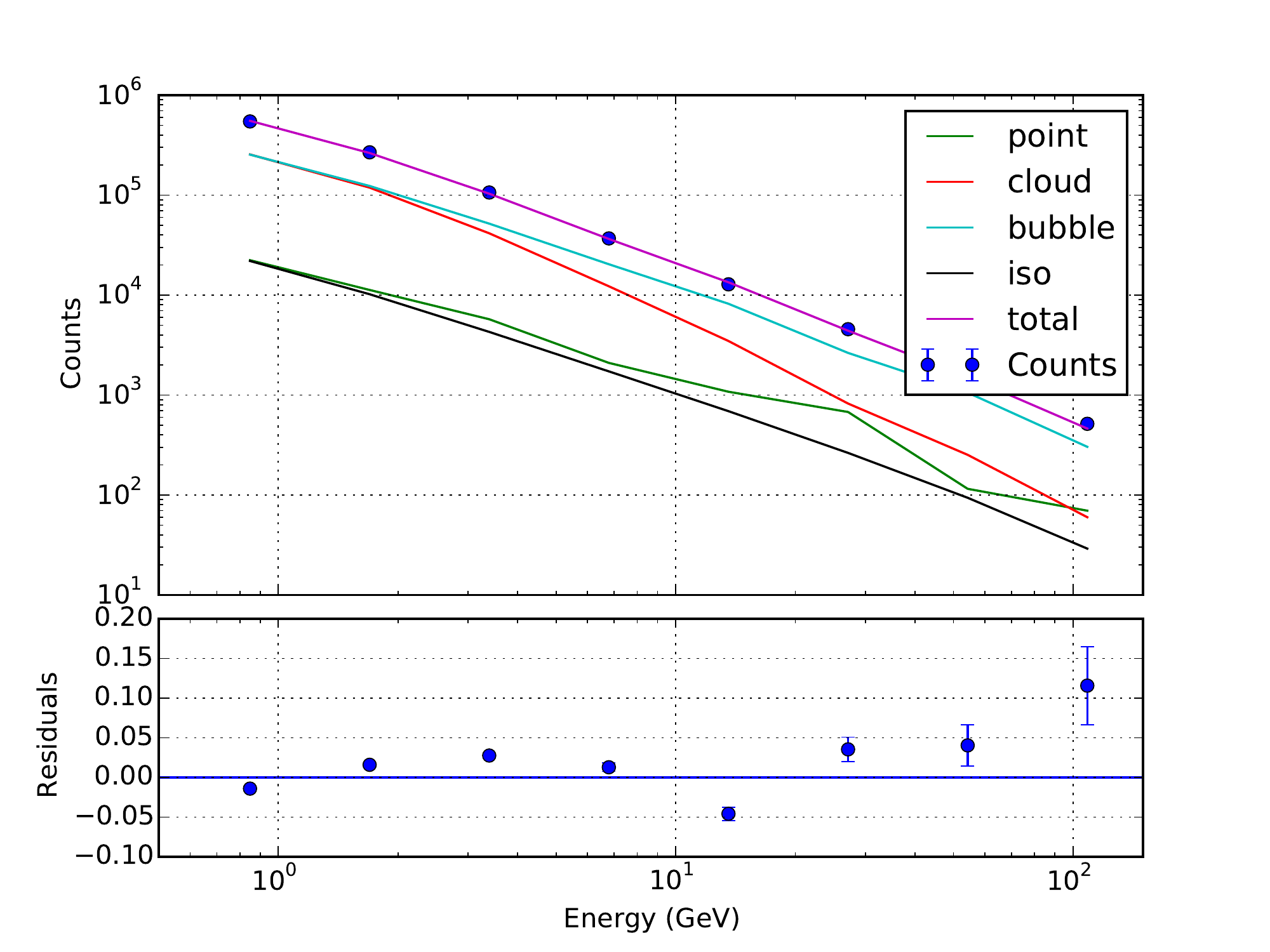}
\includegraphics[width=0.3\textwidth]{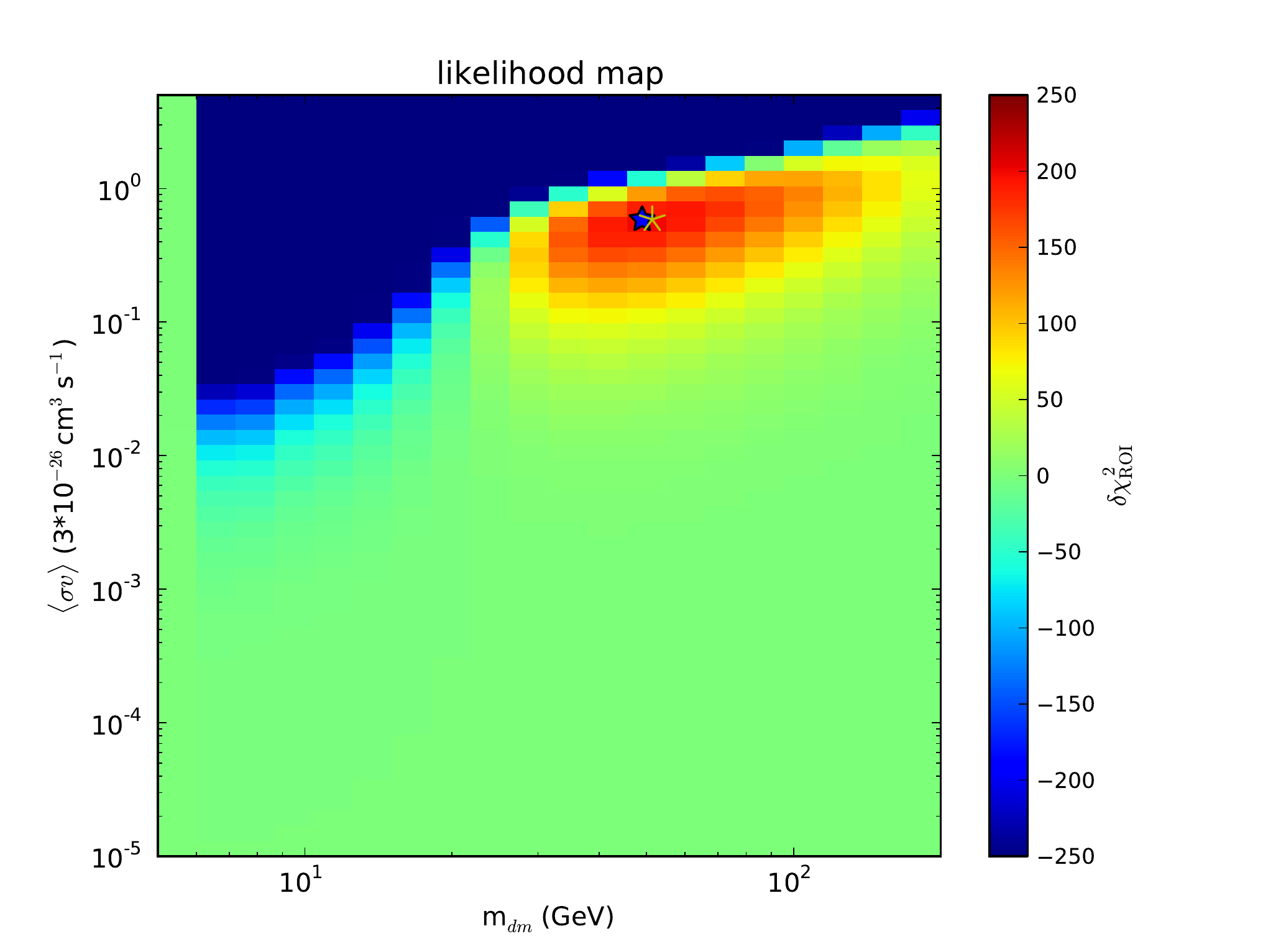}
\includegraphics[width=0.3\textwidth]{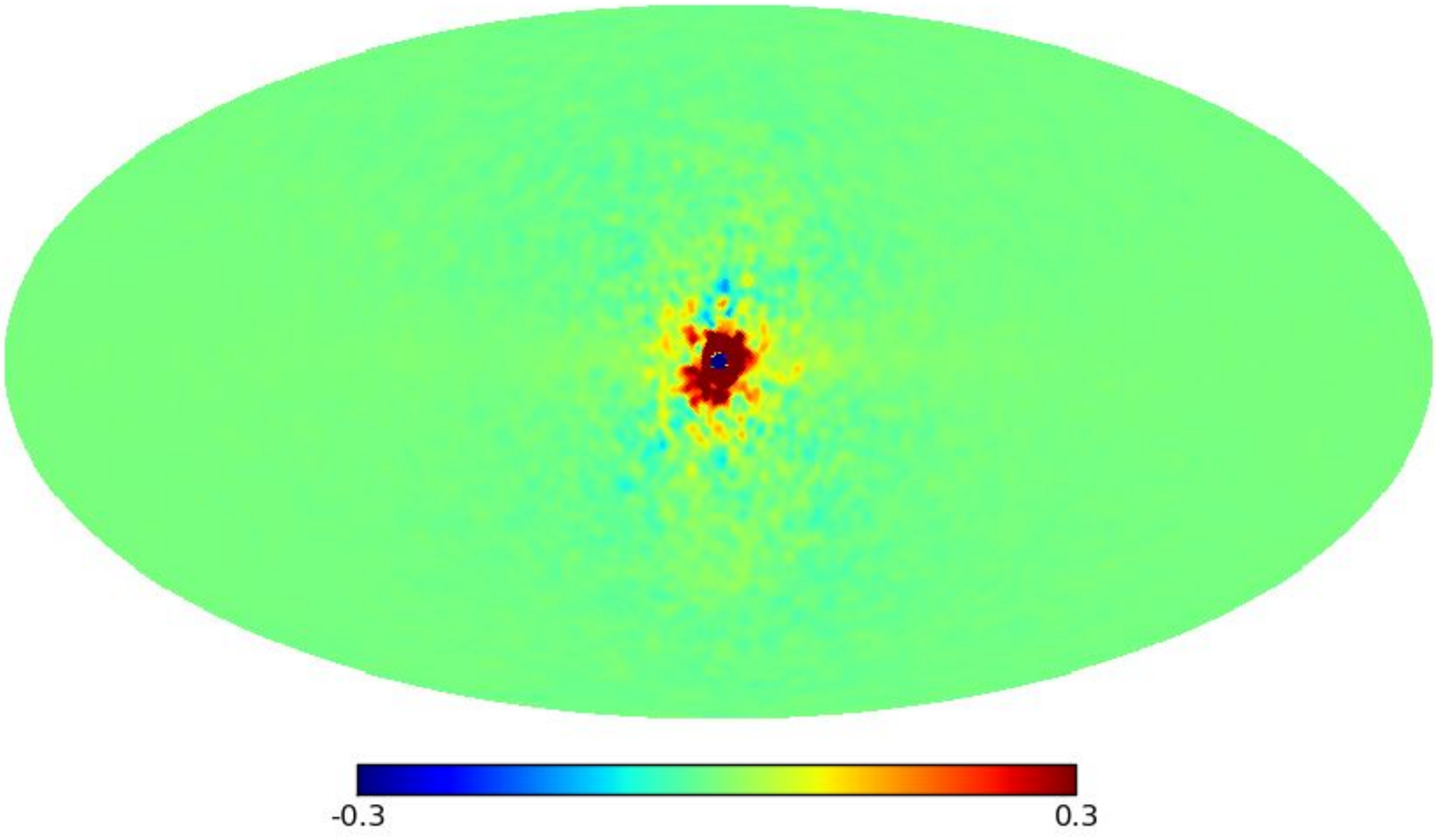}
\caption{In the left panel we show the spectral shape of different astrophysical
diffuse components after considering the isotropic emission. In the
middle panel we show the likelihood map as in Fig.~\ref{fig:likelihood map},
and in the right panel we show the pixel by pixel ts map as in Fig.~\ref{fig:pixel_by_pixel_ts}.}
\label{fig:iso}
\end{figure*}

So far,  we assumed that the DGE could be decomposed into only two
spectral components, the ``cloud-like'' and the ``bubble-like''.
At least an additional extragalactic component exist as well. This
has a different spectral form as it is dominated by active galaxies.
Its flux distribution can be approximated as being spatially isotropic
for our purpose. We investigate how much the results change when we
include this extragalactic component into the analysis. For this we
only investigate the case of the $b\bar{b}$ annihilation final states
as an illustrative example. The Fermi Collaboration did a very comprehensive
analysis of the isotropic extragalactic emission\citep{Ackermann:2014usa},
from which we adopt the spectra of the extragalactic background. This
is then converted into expected extragalactic counts $n_{iso}^{ijk}$
as described in Sect.~\ref{sub:The-likelihood}. The extended model
for the total expected counts $\lambda^{ijk}=n_{dm}^{ijk}+\alpha_{i}n_{c}^{ijk}+\beta_{i}n_{b}^{ijk}+n_{point}^{ijk}+n_{iso}^{ijk}$
is then fitted to the data in the manner described before. In Fig.~\ref{fig:iso}
the resulting spectra of the different components, the likelihood
improvement in the DM parameter plane, as well as on the sky is shown
for the primary ROI. The inclusion of this additional component does
not significantly change our results.

\section{Conclusions}

We confirm the presence of the reported apparent GC DM annihilation
signal. In contrast to previous studies, we use a phenomenological
model for the astrophysical (non-DM related) $\gamma$-ray sky, which
does not assume any spatial template and for which the spectral templates
of the astrophysical components were derived directly from the $\gamma$-ray
data. To this end, we adopted the ``cloud-like'', ``bubble-like'',
and point source components of the Fermi-LAT data analysis by Selig et al. \cite{Selig:2014qqa},
added DM annihilation flux for various DM scenarios resulting from
a NFW profile and permitted the ``cloud-like'' and ``bubble-like''
components to change their morphology to accommodate the latter. Our
best fit DM parameters are very similar to those of previous studies,
while using a different model for the astrophysical (non-DM related)
$\gamma$-ray sky. That our independent and methodologically orthogonal
modeling of the DGE confirms former findings should increase the confidence
in the existence of a GC $\gamma$-ray excess, potentially originating
from DM annihilation.

In order to investigate whether the potential DM signal is not mimicked
by a third astrophysical component, we visualized the forces acting
on the likelihood and permitted for free morphologies of the DM annihilation
signal distributions. These tests show that the data strives for a
DM component centered on the GC, as it would be expected. 
However, we find that a DM-annihilation like spectral and spatial component 
improves the fit also at locations hosting predominantly astrophysical $\gamma$-ray sources,
in particular regions hosting the hot interstellar medium. Furthermore,
regions well above the disk, which are relatively free from $\gamma$-ray
emission, are in tension with a DM-signal of the strength suggested 
by the GC region. Unless the spatial DM distribution of the Milky Way is
much less extended than described by the adopted NFW profile, the DM annihilation
interpretation of the excess flux is disfavored by the data.

These observations should be taken as a warning, indicating that also
our two component DGE model is not sufficient to capture the full
complexity of the astrophysical $\gamma$-ray sky. A third astrophysical
component seems plausible, which is potentially mimicking the GC DM-signature,
but is also present in the Galactic disk, in particular in locations
we associate with $\gamma$-ray radiation from the dilute ISM.

A better understanding of the astrophysical $\gamma$-ray radiation
is therefore necessary to confirm or refute the apparent DM annihilation
signal. Our current sensitivity is more limited by astrophysical modeling
uncertainties than by the photon count statistics. The phenomenological
and morphological investigations presented here, as well as the physical
modeling approaches by other groups, need to be refined to deal with
the large spatial and spectral complexity of the real Galactic $\gamma$-ray
emission. 

\subsection*{Acknowledgements}

We gratefully thank Simon D.~M.~White, Christoph Weniger, and an anonymous referee for feedback and comments on the manuscript. X.~H.~thanks Alejandro Ibarra for useful discussions and suggestions. T.~E.~and X.~H.~thank Andreas M\"uller and the \textit{European Centre for Theoretical Studies in Nuclear Physics and Related Areas} in Trento for the stimulating workshop atmosphere, which encouraged this research. This research was supported by the DFG cluster of excellence ʻOrigin and Structure of the Universeʼ  (www.universe-cluster.de).

\bibliographystyle{JHEP}
\bibliography{draft_gev}

\end{document}